\documentclass[twocolumn, times]{aastex63}
\usepackage{hyperref}
\usepackage{physics}
\hypersetup{colorlinks, citecolor=blue, linkcolor=blue, urlcolor=blue}
\usepackage{comment}

\newcommand{\cntext}[1]{\begin{CJK}{UTF8}{bkai}#1\ignorespacesafterend\end{CJK}} % Typeset Chinese characters
\usepackage{CJK}

\received{2022 October 4}
\revised{2023 October 27}
\accepted{2023 October 30}
\submitjournal{AAS Journals}

\shorttitle{Deep learning for particles in turbulence}
\shortauthors{Chan, Manger, et al.}
\graphicspath{{./}{./Figures/}}

\definecolor{applegreen}{rgb}{0.55, 0.71, 0.0}

\newcommand{\revOneEdit}[1]{#1}
\newcommand{\revTwoEdit}[1]{#1}    
\begin{document}
\begin{CJK*}{UTF8}{gbsn}

\title{Particle clustering in turbulence: Prediction of spatial and statistical  properties with deep learning}

%\correspondingauthor{}
%\email{}

\author[0000-0003-4469-8829]{Yan-Mong Chan}
\affiliation{Center for Computational Astrophysics, Flatiron Institute, 162 Fifth Avenue, New York, NY 10010, USA}
\affiliation{Department of Physics, The Chinese University of Hong Kong, Shatin, NT, Hong Kong}

\author[0000-0001-6552-8605]{Natascha Manger}
\affiliation{Center for Computational Astrophysics, Flatiron Institute, 162 Fifth Avenue, New York, NY 10010, USA}

\author[0000-0002-0701-1410]{Yin Li}
\affiliation{Center for Computational Astrophysics, Flatiron Institute, 162 Fifth Avenue, New York, NY 10010, USA}

\author[0000-0003-2589-5034]{Chao-Chin Yang (\cntext{楊朝欽})}
\affiliation{Department of Physics and Astronomy, University of Alabama, Box 870324, Tuscaloosa, AL 35487-0324, USA}
\affiliation{Department of Physics and Astronomy, University of Nevada, Las Vegas, 4505 S.~Maryland Parkway, Box~454002, Las Vegas, NV 89154, USA}

\author[0000-0003-3616-6822]{Zhaohuan Zhu (\cntext{朱照寰})}
\affiliation{Department of Physics and Astronomy, University of Nevada, Las Vegas, 4505 S.~Maryland Parkway, Box~454002, Las Vegas, NV 89154, USA}

\author[0000-0001-5032-1396]{Philip J. Armitage}
\affiliation{Center for Computational Astrophysics, Flatiron Institute, 162 Fifth Avenue, New York, NY 10010, USA}
\affiliation{Department of Physics and Astronomy, Stony Brook University, Stony Brook, NY 11794, USA}

\author[0000-0002-1068-160X]{Shirley Ho}
\affiliation{Center for Computational Astrophysics, Flatiron Institute, 162 Fifth Avenue, New York, NY 10010, USA}
\affiliation{Department of Astrophysical Sciences, Princeton University, Peyton Hall, Princeton, NJ 08544, USA}
\affiliation{Department of Physics, Carnegie Mellon University, Pittsburgh, PA 15213, USA}

\begin{abstract}
We investigate the utility of deep learning for modeling the clustering of particles that are aerodynamically coupled to turbulent fluids. Using a Lagrangian particle module within the Athena++ hydrodynamics code, we simulate the dynamics of particles in the Epstein drag regime within a periodic domain of isotropic forced hydrodynamic turbulence. This setup is an idealized model relevant to the collisional growth of micron to mm-sized dust particles in early stage planet formation. The simulation data are used to train a U-Net deep learning model to predict gridded three-dimensional representations of the particle density and velocity fields,  given as input the corresponding fluid fields. The trained model qualitatively captures the filamentary structure of clustered particles in a highly non-linear regime. We assess model fidelity by calculating metrics of the density field (the radial distribution function) and of the velocity field (the relative velocity and the relative radial velocity between particles). Although trained only on the spatial fields, the model predicts these statistical quantities with errors that are typically $< 10$\%. \revOneEdit{Our results suggest that, given appropriately expanded training data, deep learning could complement direct numerical simulations in predicting particle clustering within turbulent flows.}
\end{abstract}

\keywords{hydrodynamical simulations---protoplanetary disks---planet formation---neural networks}

\section{Introduction} 
\label{sec:intro}
The dynamics of aerodynamically coupled particles in turbulent fluids are of interest for astrophysical, atmospheric physics, experimental, and industrial applications \citep{toschi09}. The coupling between the fluid and particle phases modifies the turbulent cascade, and leads to clustering, effective diffusion, and collisions among the particles \citep{saffman56}. The generic problem has numerous physically significant variations, depending upon the regime of aerodynamic drag (Stokes or Epstein), the compressibility of the turbulence, the relative mass in particles versus that in the fluid, and the amount of fluid volume excluded by particles' finite size.

In the astrophysical domain, the response of solid particles to turbulence is important in protoplanetary disk physics and early phase planet formation. On large spatial scales, the ubiquitous substructure seen in dust continuum images of protoplanetary disks \citep{vdm13,alma14,andrews18,andrews20} is interpreted as concentrations of solids within gaseous vortices \citep{barge95,bracco1999particle} and axisymmetric pressure maxima \citep{whipple72}, structures which may themselves have formed as a consequence of disk turbulence \citep{johansen09,manger18}. On small scales, the collision velocity of small (micron to cm-sized) particles is typically determined by how the particles interact with turbulence in the gas \citep{weidenschilling93,johansen14}. For small particles the collision velocity induced by turbulence increases with size. Because dust particles fragment (or bounce) above some threshold velocity \citep{blum08}, this behavior leads to the prediction of a {\em maximum particle size} that can result from coagulation. On intermediate scales, the two-way momentum transfer between particles and gas in disks leads to the {\em streaming instability} \citep{youdin05,krapp19,paardekooper20,yang21}, that may be responsible for the formation of planetesimals via gravitational collapse \citep{johansen15,simon16,li21,yang17,schafer17}.

The various manifestations of particle-gas interactions in protoplanetary disks can be studied using analytic methods \citep[e.g.][]{volk80,ormel07,pan10,krapp19,lin21}, and by simulations that treat the particle phase as either discrete Lagrangian particles (e.g. \citet{yang16}) or as a second fluid \citep[e.g.][]{bracco1999particle,pan11,ishihara18,bhatnagar18,krapp20,sakurai21,huang2022multifluid}. Although powerful, these methods do not suffice to answer all of the physically interesting questions that arise in planet formation studies. Numerical simulations can only model a small subset of the full range of spatial scales, leaving open questions as to how interactions on different scales combine \citep{chambers10,hartlep20}. Even when only a limited range of scales need to be studied, it is computationally expensive to run models that span the range of plausible size distributions for particles in protoplanetary disks. Machine Learning (ML) approaches may provide new theoretical tools to study these problems. In analogous cosmology applications, ML has been used to generate realistic predictions of gas properties from dark matter only simulations, and to interpolate reliably from grids of simulations that sparsely cover the full parameter space \citep{vn21}. \revOneEdit{ML architectures have also been combined with traditional turbulence modeling techniques, such as Reynolds-averaged Navier-Stokes (RANS) models and large eddies simulation (LES) models, to improve models of the turbulence properties and turbulence-particle interactions predicted by coarse-scale, non-ML models \citep{ling2016reynolds,xie2019modeling,wang2017physics,wu2022large,davydzenka_tahmasebi_2022}. These ML methods are able to recover the statistical properties of turbulence and particle clustering over a reasonable range in parameter space of interest.} 
%\timoedit{[TIMO: Maybe we can add a paragraph explaining/ describing why LES etc. are not popular in planetary situations? This can actually be a good point to justify the use of our model.]}
In planet formation it might be possible to extend the range of explicit simulations with an effective ML sub-grid model, and to make predictions for arbitrary particle size distributions using a fairly small set of input simulations. 

In this paper we present an initial investigation into how well established ML methods work for predicting the spatial and statistical properties of particles in turbulent fluids. We consider the response of a single particle size to isotropic forced turbulence, in a periodic domain, with aerodynamic drag in the Epstein regime. This setup excludes the forces that lead radial and vertical epicyclic oscillations of free particles, which are physically important \citep{youdin07}. It is, however,  the simplest model problem that is relevant to studies of the rate and velocity of particle collisions on small scales in protoplanetary disks. Clustering in this regime is well-studied, and is understood to result from the interaction of particles with small-scale fluid vortices \citep[e.g.][]{johansen14}. This physical understanding does not, however, lead to a predictive model for the particle distribution. The collision velocity, on the other hand, is amenable to an analytic model \citep[several exist, the most common in disk applications being][]{ormel07}. We generate training data using a recently developed Lagrangian particle module within the {\tt Athena++} code (Yang et. al., in prep.), and use a U-Net network architecture \citep{ronneberger2015u} to learn the mapping between the three-dimensional fluid fields and gridded representations of the particle data. We then assess the fidelity of the predictions for both statistical properties of the particles (the strength of clustering and the relative velocity as a function of scale), and for their instantaneous spatial distribution. 

The plan of this paper is as follows. In \S\ref{sec:simulations} we describe the {\tt Athena++} code, the implementation of Lagrangian particles within the code, and the setup of the forced turbulence simulations. The Machine Learning methods are described in \S\ref{sec:ML}. Sections \S\ref{sec:density} and  \S\ref{sec:normal_particle_velc_prediction} quantify the network performance for the particle density and velocity fields, and for statistical properties that would be relevant to the calculation of the rate and mutual velocity of particle collisions. We present some generalization tests in \S\ref{sec:generalization} and conclude in \S\ref{sec:conclusion}.

%rename to generation of training data? or something more general?
\section{Numerical simulations} 
\label{sec:simulations}

%moved before description of simulations as necessary quantities are introduced here.
\subsection{Physics of turbulent clustering}
\label{sec:phy_of_clustering}
To aid in the subsequent discussion of clustering statistics, we briefly introduce some relevant physics of turbulence and turbulence-driven dust dynamics.

In the standard description of fluid turbulence \citep{kolmogorov41,kolmogorov1991local}, the  statistics of turbulence within the inertial range depend only on the mean energy transfer rate $\langle \epsilon \rangle$. In this limit, the two point relative velocity $\langle \delta \vec{u} \cdot \delta\vec{u}\rangle$ at separation $\ell$ scales as $\langle \epsilon \rangle^{2/3} \ell^{2/3}$. In the inertial range, dissipation due to viscosity is negligible and energy cascades to smaller scales. The scale at which viscosity becomes important can be found by equating $\langle \epsilon \rangle $ to the energy dissipation rate $\nu \left[\delta u(\ell_d)/\ell_d\right]^2 \approx \langle \epsilon \rangle$ \citep{ching2014statistics}, from which one obtains the
Kolmogorov dissipative length scale, timescale and velocity scale as $\ell_\eta = (\nu^3/\langle \epsilon \rangle)^{1/4}, \tau_\eta = (\nu/\langle \epsilon \rangle)^{1/2}$, and $u_\eta = (\nu \epsilon)^{1/4}$ respectively. 

Now consider dust particles coupled to such a turbulent flow. If friction is the dominant force driving the dust, the equation of motion of a dust particle takes the form,
\begin{equation}
    \frac{d\vec{v}_p}{dt} = \frac{\vec{v}_g(p) - \vec{v}_p}{\tau_s},
    \label{eq:particle_eom}
\end{equation}
where $\vec{v}_p$ is the velocity of particles, $\vec{v}_g$ is the velocity of gas, and $\tau_s$ is the friction time characterizing the strength of coupling between the dust and gas. For dust particles in the Epstein regime\footnote{In this regime the drag force is {\em linear} in the velocity difference between the particle and the fluid, rather than quadratic as in the Stokes regime typically relevant to terrestrial fluids}, that is, with size $a_p$ much smaller than the mean free path $\lambda_g$ of the gas molecules $(a_p\ll \lambda_g)$, the friction time is given by \citep{weidenschilling93},

\begin{equation}
    \tau_s = \left(\frac{\rho_d}{\rho_g} \right) \left(\frac{a_p}{c_s}\right),
    \label{eq:epstein_drag}
\end{equation}
where $\rho_d$ is the material density of the dust, $\rho_g$ is the fluid density, and $c_s$ is the local speed of sound of the gas.

The clustering behavior depends on how $\tau_s$ compares with a characteristic time-scale of the turbulent flow. A commonly used time scale relevant for our purposes is the Kolmogorov time-scale $\tau_\eta$ \citep{sundaram_collins_1997,pan11,ishihara18}. It represents the turnover time of the smallest eddies. The ratio of these time-scales is the dimensionless Stokes number $\text{St} = \tau_s/\tau_\eta$\footnote{In protoplanetary disk studies, the ``Stokes number" is often defined instead using as a characteristic time-scale the dynamical time, $\Omega_{\rm K}^{-1}$, where $\Omega_{\rm K}$ is the Keplerian orbital frequency. Because our simulations do not include shear, we use the definition appropriate for fluid turbulence.}. Particles with $\text{St} \ll 1$ are well coupled to the gas at all scales and behave like trace particles. With very little memory of its previous motion, the clustering will be mainly determined by the velocity divergence of the surrounding flow at one point in time. It has been shown that the particle velocity divergence $\sum_i \partial_i v_{p}^i \approx  \text{St}/\tau_\eta$ \citep{pan11}, where $i$ is the spatial index. This implies that for $\text{St} < 1$ the clustering intensifies with Stokes number. 

Conversely, when $\text{St} > 1$, turbulent fluctuations are too rapid for the particles to respond. This has two effects. First, as the friction time $\tau_s$ increases, the history of the particle has an increasingly large influence on the particle relative velocity. This can be seen in the formal solution of equation~(\ref{eq:particle_eom}) \citep{pan2013turbulence},
\begin{equation}
    \vec{v}_p(t) = \frac{1}{\tau_s} \int_{t_0}^{t} \vec{v}_g(\mathbf{X}(\tau),\tau) \exp(-\frac{t-\tau}{\tau_s}) d\tau, 
\end{equation}
where $\mathbf{X}(t)$ is the position of the particle at some time $t$ and $t_0$ is a reference time chosen to be long enough (i.e. $t-t_0\gg \tau_s$) so that the velocity at time $t_0$ has been ``forgotten" by the particle. This solution implies that the particle's current velocity depends on the past history of the gas velocity within a time window of $\approx \tau_s$. Hence, one should expect that for $\text{St} > 1$ the local gas velocity divergence, and also the $\text{St}/\tau_\eta$ scaling discussed above, should no longer be adequate for describing particle clustering. 

Furthermore, particles with $\text{St} > 1$ respond to eddies of different turnover time selectively, giving rise to a characteristic clustering scale $\ell_{\tau_s}$ \citep{pan11}. Since $\tau_s > \tau_\eta$, some of the eddies have turnover time smaller than $\tau_s$. Particles at these scale are decoupled from the flow and experience random kicks due to the flow. This inhibits clustering at those scales. On the other hand, particles with turnover time much longer than $\tau_s$ are well coupled to the eddies. This depresses clustering at large scale. Hence, the most significant clustering is expected to occur when $\tau_{eddy} \approx \tau_{s}$. In the inertial range, $\tau_{eddy} \approx \ell^{2/3} \langle \epsilon \rangle^{-1/3}$, where $\ell$ is the characteristic size of the eddies and $\langle \epsilon\rangle$ is the mean energy injection rate. Equating $\tau_{eddy}$ with $\tau_s$ allow us to obtain the characteristic scale of the clustering $\ell_{\tau_s} \approx \langle \epsilon \rangle^{1/2} \tau_s^{3/2} = \text{St}^{3/2} \ell_\eta$, where $\ell_\eta$ is the Kolmogorov length.

\subsection{Forced turbulence simulations}
\label{sec:forced_turbulence_simulation}
We use the \texttt{ATHENA++} magnetohydrodynamics (MHD) code \citep{stone20} to calculate the training data for our study. \revOneEdit{The fluid system solved represents the equations of compressible, inviscid hydrodynamics with an isothermal equation of state,
\begin{eqnarray}
  \frac{\partial \rho_{\rm gas}}{\partial t} + \nabla \cdot \left( \rho_{\rm gas} {\bf v} \right) & = & 0, \\
  \frac{\partial \rho_{\rm gas} {\bf v}}{\partial t} + \nabla \cdot \left( \rho_{\rm gas} {\bf v}{\bf v} + {\bf P}^* \right) & = & 0, \\
  P_{\rm gas} & = & \rho_{\rm gas} c_s^2. 
\end{eqnarray}
Here, ${\bf P}^*$ is a diagonal tensor with components $P_{\rm gas}$, and $c_s$ is the constant speed of sound. In common with \citet{pan11}, who also used a Godunov-type numerical scheme, the simulations do not include any explicit viscosity (either a Navier-Stokes viscosity or a hyperviscosity). Dissipation of kinetic energy occurs via numerical diffusion at the grid scale. This choice maximizes the extent of the inertial range in the simulated turbulence -- important given the moderate resolution we are using -- but means that the effective viscosity and corresponding Reynolds number can only be estimated using established scaling relations for turbulence \citep{pan11}.}
All simulations are conducted using a uniform 3D Cartesian grid for the gas variables with a resolution of $256$ cells in each spatial direction.
We use the Harten--Lax--van Leer--Einfeld (HLLE) Riemann solver combined with a second-order accurate van Leer predictor-corrector scheme for time integration. Spatial reconstruction is done using the 3rd order piecewise parabolic method (PPM). We use periodic boundary conditions in all directions.

The gas is initialized with constant density $\rho_\mathrm{gas} = 1$ and with all velocity components $v_i =0$. The speed of sound $c_\mathrm{s}=1$.
To generate a turbulent velocity spectrum for the gas, we use the FFT turbulence driver module of \texttt{ATHENA++}, which continually feeds energy into the domain at the largest scales by adding large scale velocity perturbations to the flow. The scale of these perturbations is controlled by the parameters $k_\mathrm{min}$ and $k_\mathrm{max}$, denoting the smallest and largest wavenumber to be perturbed. (Here wavenumber is defined as $k=1/\lambda$, without a factor of $2\pi$). Unless otherwise specified, we use $k_\mathrm{min}=1$ and $k_\mathrm{max}=5$ with simulation box size $L_x=L_y=L_z=1$, which allows the turbulent energy spectrum to develop naturally into a \revTwoEdit{Kolmogorov-like spectrum, with numerical dissipation occurring near the Nyquist frequency. We note that, due to the limited numerical resolution of the training simulations, the obtained energy spectrum of the turbulence is not a standard Kolmogorov type spectrum.} We exclude data from an initial transient state from our ML dataset. The amount of energy injected per perturbation interval (which we set to be equal to one simulation time-step) is given by the energy injection rate $\epsilon_\mathrm{inj}$ and is set to $\epsilon_\mathrm{inj} = 0.005$ (except where specified). The RMS gas velocity calculated from the simulation is $v_{g}^\text{rms} \approx 0.104$, which is $\approx 10\%$ the speed of sound. The turbulence, although computed with a compressible code, is thus, for practical purposes, almost incompressible. As we are investigating subsonic turbulence, we enforce solenoidal turbulence driving by suppressing the compressive component of the velocity perturbation.

\subsection{Lagrangian dust particles}
\label{sec:lagrangian_dust_particles}
We use the Lagrangian dust particle module of the \texttt{ATHENA++} code (Yang et. al., in prep.) to simulate the particle dynamics. In all our simulations, $256^3$ dust particles are used.
For simplicity, we only treat the drag force exerted on the dust particles by the gas and neglect the back-reaction of the particles on the gas, which is a good approximation in cases with low dust-to-gas mass ratios. To interpolate between the gridded values of the gas component and the particle positions, the module uses the triangular shaped cloud (TSC) algorithm. 
Unless otherwise specified, we use a particle friction time of $\tau_s = 0.1$.

From the simulated output, we estimate the turbulent parameters of the fiducial test case to be: $\langle \epsilon\rangle\approx 3.17\times 10^{-3}$, $\nu \approx 1.95 \times 10^{-5}$, $\ell_\eta \approx 1.24\times 10^{-3}$, $\tau_\eta \approx 7.84\times 10^{-2}$ and $u_\eta \approx 1.57\times 10^{-2}$, all in the relevant simulation units. \revOneEdit{The macroscopic Reynolds number of the fluid simulation is computed to be $\text{Re} \approx 3400$.} The particle clustering length is estimated to be $\ell_{\tau_s} \approx 1.58\times 10^{-3}$. The energy injection rate $\langle \epsilon\rangle$ and eddy turnover time $\tau_\eta$ are estimated from the third-order structure function and strain-rate tensor respectively, the particle friction time is taken directly from simulation setup, and the rest are obtained from their relationships with these three variables. 
\section{Machine learning methods}
\label{sec:ML}

\subsection{Dataset}
Snapshots of the simulation described above are used as the training set of the model. The temporal separation between the frames is chosen to be long enough so that adjacent frames are uncorrelated. The gas density field $\rho_g$, gas velocity field $\vec{v}_g$, particle density field $\rho_p$, and particle velocity field $\vec{v}_p$ are used to train and validate the network. All fields are gridded on a $256^3$ mesh. (The gas fields are natively gridded; the particle data is mapped to the same grid). The first two serve as the model's inputs, and the latter as outputs. 
\revOneEdit{The particles are gridded because our network architecture and data augmentation techniques (See \S\ref{sec:ml_model}, \S\ref{sec:ml_training} for details) work on gridded inputs and outputs only.} 

%\mledit{For example, a common method of producing lagrangian particle outputs directly is to output a $\mathbb{R}^{N\times(3+3)}$ vector representing the positions and velocities of the $N$ particles \citep{ummenhofer2020lagrangian,zhang2020fluidsnet}. For this type of architecture, the number of particles is hard-coded into the model. However, this is incompatible with our training strategy. In our work, the input fields are cropped into smaller pieces to make it easier for the network to learn the translational and rotational invariance of the flow (see \S\ref{sec:ml_training} for details). As the number of particles in the cropped pieces $M$ can be variable, this encoding does not work in our case. (Ok this might not be good, because people have opt toward using rough simulation and let the ML to improve the forcing prediction)}

%\mledit{We choose to grid the particles because our network architecture - the U-Net (See \S~\ref{sec:ml_model} for the details) involves convoluting the input fields with fixed sized kernels, which most straight forward to implement on ordered and regular grids. Furthermore, although gridding will inevitably remove small scale information from our datasets, application of similar methods in cosmology to predict corrections for N-body simulations \citep{jamieson2022field} shows . Therefore, as preliminary test of applying deep learning on the problem of particle clustering, we choose train our network with it} 
\subsection{Data preprocessing and Normalization}
Data preprocessing and normalization steps are applied to the input and output fields to aid the training.
\subsubsection{Input fields}
The gas density $\rho_g$ is log normalized,
\begin{equation}
    \rho_g^\ast = \log(\rho_g)
    \label{eq:rhog_norm},
\end{equation}
where $\rho_g^\ast$ is the normalized gas density and $\rho_g$ is the gas density. % and $\epsilon$ is a small factor taken to be $10^{-8}$ to prevent taking logarithm of zero. %Strictly speaking this is unnecessary because $\rho_g$ is close to unity but we include it because it is in the code. 
The log normalized density field helps separate gas density of various scales, allowing the network to better perceive the input field. No preprocessing is applied to the gas velocity $v_g$. \revOneEdit{In preliminary experiments, we considered preprocessing the velocity field to show the network the gas vorticity, based on the physical argument that particle clustering depends on how the particles respond to vortices of varying scale. Somewhat surprisingly, we found that a network trained using gas vorticity $\nabla \times \vec{v}_g$ as input produces visually indistinguishable outputs from one trained directly on gas velocity. While a qualitative study of the relevant statistics has not been performed on these models, we believe the visual comparison shows that the network is powerful enough to itself extract the physically relevant quantities from the velocity field. We will therefore not consider networks trained with vorticity (or velocity divergence) inputs in the remainder of the paper.}

%\mledit{The gas velocity is chosen as the input field because it is easy for the network to extract physically relevant quantities such as vorticity and velocity gradients from the velocity field. Indeed, we found that a network trained using gas vorticity $\nabla \times \vec{v}_g$ as input produces visually indistinguishable output from that trained on gas velocity. While a qualitative study of the relevant statistics has not been performed on these models, we believe the visual comparison is sufficient to justify our choice of the input fields. We will not consider networks trained on velocity divergence and vorticity inputs in the remainder of the text.}

\subsubsection{Output fields}
The particle density field $\rho_p$ and velocity field $v_p$ are non-gaussian \citep[e.g.][]{ishihara2018dust}. Therefore, data normalization and preprocessing techniques are employed to aid the training of the neural network.

For the particle density field $\rho_p$ we apply the normalization,
\begin{equation}
    \rho^\ast_p = \ln(\exp(\rho_p/\rho_0)-1).
    \label{eq:rhop_norm}
\end{equation}
Here, $\rho^\ast_p$ is the normalized particle density and $\rho_0=0.14$, which corresponds roughly to the average particle density, is empirically chosen to make the distribution of normalized density $\rho^\ast_p$ most symmetric. The mapping between $\rho_p^\ast$ and $\rho_0$ is one-to-one, allowing us to invert $\rho_p^\ast$ using a softplus-like function,
\begin{equation}
    \rho_p = \rho_0\ln(\exp(\rho_p^\ast)+1).
    \label{eq:rhop_inv_norm}
\end{equation}
For the particle velocity field $\vec v_p$, the following normalization is applied,
\begin{equation}
    \vec{v}_{rel} = \vec{v}_p - \vec{v}_g,
    \label{eq:vp_norm}
\end{equation}
where $\vec v_g$ is the gas velocity, $\vec v_p$ is the particle velocity, and $\vec v_{rel}$ is the relative velocity. 
The motivation for this is to try to prevent the bulk gas velocity $\vec v_g$ from masking the fine details of the particle velocity $\vec v_p$ by training the network (referred to later as UNET-R) on the relative velocity $\vec v_{rel}$. Intuitively, this helps the network focus on the physically interesting differences between the particle and gas velocities, instead of learning to copy the background flow. To evaluate the effectiveness of this strategy, we also trained a network directly on $\vec v_p$ (later, UNET-V) and compared it against that trained on $\vec v_{rel}$. Details of the comparison are presented in \S\ref{sec:normal_particle_velc_prediction}.

\subsection{Model and training}

\subsubsection{Model}
\label{sec:ml_model}
\begin{figure*}
    \centering
    \includegraphics[width=0.9\textwidth]{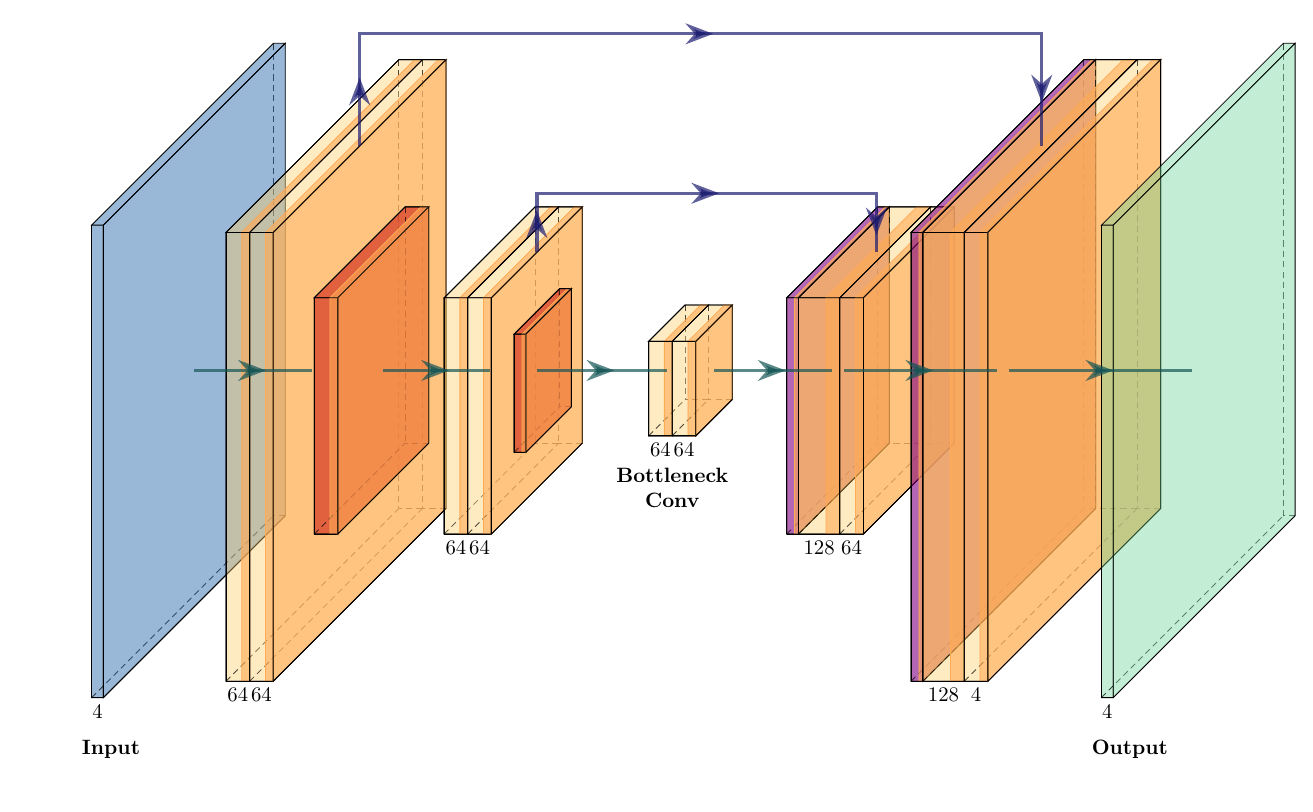}
    \caption{The architecture of the U-Net. The thickness of individual layers corresponds roughly to the number of channels in each layer and the width and height corresponds to the spatial size. The color scheme corresponds to: convolution layers of kernel size 3 and stride 1 \textbf{(orange)}, Leaky ReLU activation with batch normalization \textbf{(deep orange)}, downsampling layers using convolution of kernel size 2 and stride 2 \textbf{(red)} and upsampling layers using transposed convolution of kernel size 2 and stride 2 \textbf{(purple)}. Residue connection is implemented for all convolution layers except those used in up and down sampling. Similar to other U-Net architectures, large scale information is incorporated by narrow down and concatenating compression layer outputs with the upsampled outputs.} 
    \label{fig:unet_architecture}
\end{figure*}

The network architecture U-Net, first developed by \cite{ronneberger2015u} for biomedical image segmentation and later generalized by \cite{milletari2016v} to handle volumetric data, is used. U-Net is a ``fully convolutional network" that is capable of taking a multichannel, 3D input block and producing an output block of similar spatial size but with a different number of channels. It exhibits translational equivariance -- that is, its predictions depend only on the relative positions of the grid points \citep{long2015fully}. This is desirable because it is consistent with the physical picture of homogeneous turbulence \citep{ching2014statistics} and because it allows the network to be applied to input fields of different spatial sizes. The U-Net has been successfully employed in cosmology for tasks including super-resolution (generating high-resolution predictions from low-resolution simulations) \citep{schaurecker2021super}, and prediction of galaxy distributions from the dark matter density field \citep{yip2019dark}. %This motivates us to apply the architecture to our problem.

The network is implemented using the \textbf{map2map} neural network emulator with the architecture specified in Fig.~\ref{fig:unet_architecture}. The network consists of two major parts, the ``compression part" and the ``decompression part". The former consists of successive convolutional and downsampling layers that compress information of the input field into a smaller size latent representation, while the latter consists of successive levels of convolutional and up-sampling layers that expand the latent information into the output block. Information extracted in different levels of the network is incorporated by concatenating outputs of the compression part and those of the expanding part on the same level. 

\subsubsection{Training} 
\label{sec:ml_training}

The simulation frames, 252 in total, are randomly divided into training and validation sets in a 4:1 ratio. Training is done using only the frames in the training set. The entire training set is iterated through during each epoch of training. \revOneEdit{During training, for each frame in the train set, a $128^3$ block is cropped from the input fields at 8 floating anchors\footnote{The eight anchors are chosen to be the upper-right corners of the eight octants of an input block. The anchors are floating because, during training (not in validation and output generation), they are randomly displaced by at most four grid points along each spatial direction so that each time the networks see a slightly translated version of the input. }. Each $128^3$ block is padded $20$ grids in all directions to produce input fields of dimension $(4,168,168,168)$, with three of the input channels coming from the spatial components of the gas velocity $v_g^{(x/y/z)}$ and one from the gas density $\rho_g$. During both cropping and padding, periodic boundary conditions are applied when the cropping or padding array index exceeds the valid range. Discrete rotation and parity transforms are also applied to the input block using map2map. }

\revOneEdit{The data augmentation process serves several purposes. Practically, the cropping allows the input to fit onto one GPU, which enhances parallelization. Physically, cropping the data about floating anchors and applying rotation and parity flips helps the network capture the turbulent flow's transitionally symmetric and isotropic nature. Finally, the padding helps the network to preserve complete translational symmetry across the boundary and removes boundary artifacts in the output fields \citep{schaurecker2021super}}.  %we divide the input fields into eight octants of size $128^3$ so that they can fit onto one GPU.  Then, we pad them periodically by $5$ grids on all sides so that they are $148^3$ in size. The padding, as discussed by \cite{schaurecker2021super}, allows the network to preserve full translational symmetry and generate outputs without boundary artifacts.

 The input enters the U-Net and produces an output field of size $(4,128,128,128)$: One of the channels represents particle density $\rho_p$, and the rest represent particle relative velocity $\vec{v}_{rel}$. Finally, the whole $(4,256,256,256)$ field is reconstructed from the output patches produced by map2map.

For the loss function, we choose to minimize the mean square error (MSE) loss over all output channels. Weight decay is also used. The loss function, which is minimized using the Adam optimizer \citep{kingma2014adam}, takes the following form,
\begin{equation}
    \mathcal{L} = \left[\frac{1}{\mathcal{B}}\sum_{i=1}^{\mathcal{B}}(\mathbf{y}_\text{tgt} - \mathbf{y}_\text{out}^{(i)})^2\right] + \lambda \norm{\mathbf{w}}^2,
\end{equation}
where $\mathbf{y}_\text{tgt},\mathbf{y}^{(i)}_\text{out}$ represents the target and output fields, $\mathcal{B}$ is the batch size, $\lambda$ is the weight decay, and $\norm{\mathbf{w}}$ is the square norm of weights. Unless otherwise specified, all training runs in this paper use learning rate $10^{-4}$, weight decay $\lambda= 10^{-4}$, and batch size $\mathcal{B}=1$, for around 400 epochs until the results are stable.

After training, we evaluate the model by comparing the output with the target over a variety of statistics. The results are summarized in \S\ref{sec:density} and  \ref{sec:normal_particle_velc_prediction}.

\section{Prediction of particle density field}
\label{sec:density}

\subsection{Visual Comparison}
\label{sec:rhop_visual_comp}
\begin{figure*}
    \centering
    \includegraphics[width=\textwidth]{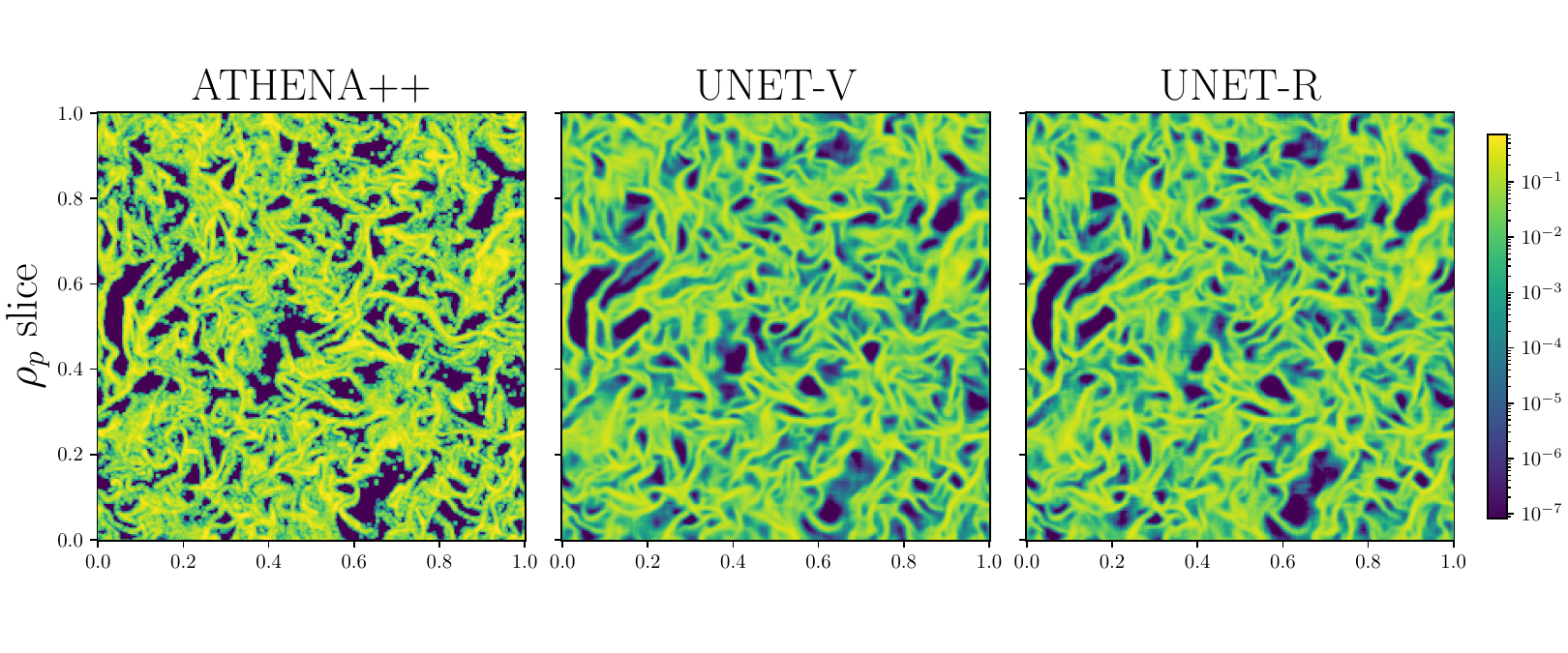}
    \vskip-0.75cm
    \caption{Particle density field slice $\rho_p$ from ATHENA++ (left), UNET-V (center, a network trained on the relative velocity between particles and gas) and UNET-R (right, a network trained on the absolute particle velocity), for a typical frame. Qualitatively, there is good agreement between the predicted structures and the ground truth, though the network outputs are visually less sharp than the target simulation data.}
    \label{fig:rhop_slice_comp_normal}
\end{figure*}

\begin{figure}
    \centering
    \includegraphics[width=0.4\textwidth]{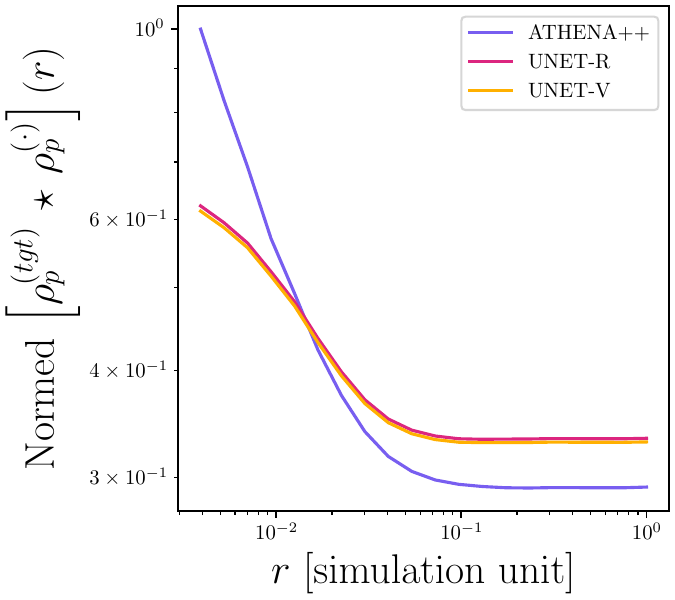}
    \caption{Normalized cross correlation of $\rho_p$ of target field (ATHENA++) with itself, UNET-R and UNET-V. The peak values of the statistics are: 1.0, 0.622 and 0.613 respectively.}
    \label{fig:rhop_cross_corr}
\end{figure}

Using the trained network, we output the predicted particle density field $\rho_p$ and compare it to the target on a frame-by-frame basis. Snapshots of selected frames are presented in Fig.~\ref{fig:rhop_slice_comp_normal}. Because the densities span a wide range, a logarithmic color map is chosen to aid visualization. We also show the normalized cross correlations\footnote{The normalized cross correlation of two function $f,g$ is defined to be $\left[(f\star g)(r)\right]/\sqrt{\left[(f\star f)(0)\right]\left[(g\star g)(0)\right]}$, which is a normalized inner product of $f,g$. For volumetric data, the result is radially binned.} of these frames, defined as in Fig.~\ref{fig:rhop_cross_corr}. 

As shown in Fig.~\ref{fig:rhop_slice_comp_normal}, both machine learning outputs (UNET-V, UNET-R) produce density distributions that are similar to the ground truth (ATHENA++). Locations with voids in the density distribution align, and the filamentary structures of the dust are captured. However, both predicted outputs are less sharp than the target field, indicating that the algorithm has smoothed out the densities. The density at the locations of peaks is also less pronounced, indicating an under-prediction in those areas. 
These observations are borne out in Fig.~\ref{fig:rhop_cross_corr}, where we see that the peak values of normalized cross correlations for both UNET-V and UNET-R are only $\approx 0.6$. The larger peak values for UNET-R show that UNET-R, trained on the relative velocity, is slightly better at predicting $\rho_p$ than UNET-V.

\subsection{Radial Distribution Function (RDF)}
\label{sec:normal_stat_rdf}
We compute the particle radial distribution function (RDF) $g(r)$ of the simulated and U-Net-produced outputs to study the difference in their spatial distributions more carefully. Let $\bar{n}$ be the average dust density. The RDF is defined so that the average number of dust particles $P(r)$ inside a volume element $dV$ at a distance $r$ relative to a reference particle is given by,
\begin{equation}
    P(r) = \bar{n} g(r) dV. 
    \label{eq:radial_distribution_func}
\end{equation}
Although this definition can be applied directly to compute the RDF, a faster way of obtaining it on a grid is available. First, we compute the particle density correlation function (DCF) $\xi(r)$, which is defined as,
\begin{equation}
    \xi(r) = \frac{1}{\bar{n}^2} \langle \left(n(\vec{x})-\bar{n}\right) \left(n(\vec{x}+\vec{r})-\bar{n}\right)\rangle .
    \label{eq:density_correlation_func}
\end{equation}
Then we use the relation $g(r) = 1 + \xi(r)$ between the two quantities to obtain the RDF \citep{shaw2003particle}. This allows for an efficient computation, as  Eq.~\ref{eq:density_correlation_func} can be computed quickly using the fast fourier transform (FFT). Unless otherwise specified, all RDFs computed in this paper are based on this method. The use of grid-based metrics, rather than metrics computed directly from the particle distribution, differs from most prior work. Differences in results that arise between the method based on Eq.~\ref{eq:radial_distribution_func}, and that based on Eq.~\ref{eq:density_correlation_func} for the simulated particle density, are discussed in Appendix~\ref{sec:grid_stats_vs_part_stats}.

\begin{figure*}
    \centering
    \includegraphics[width=\textwidth]{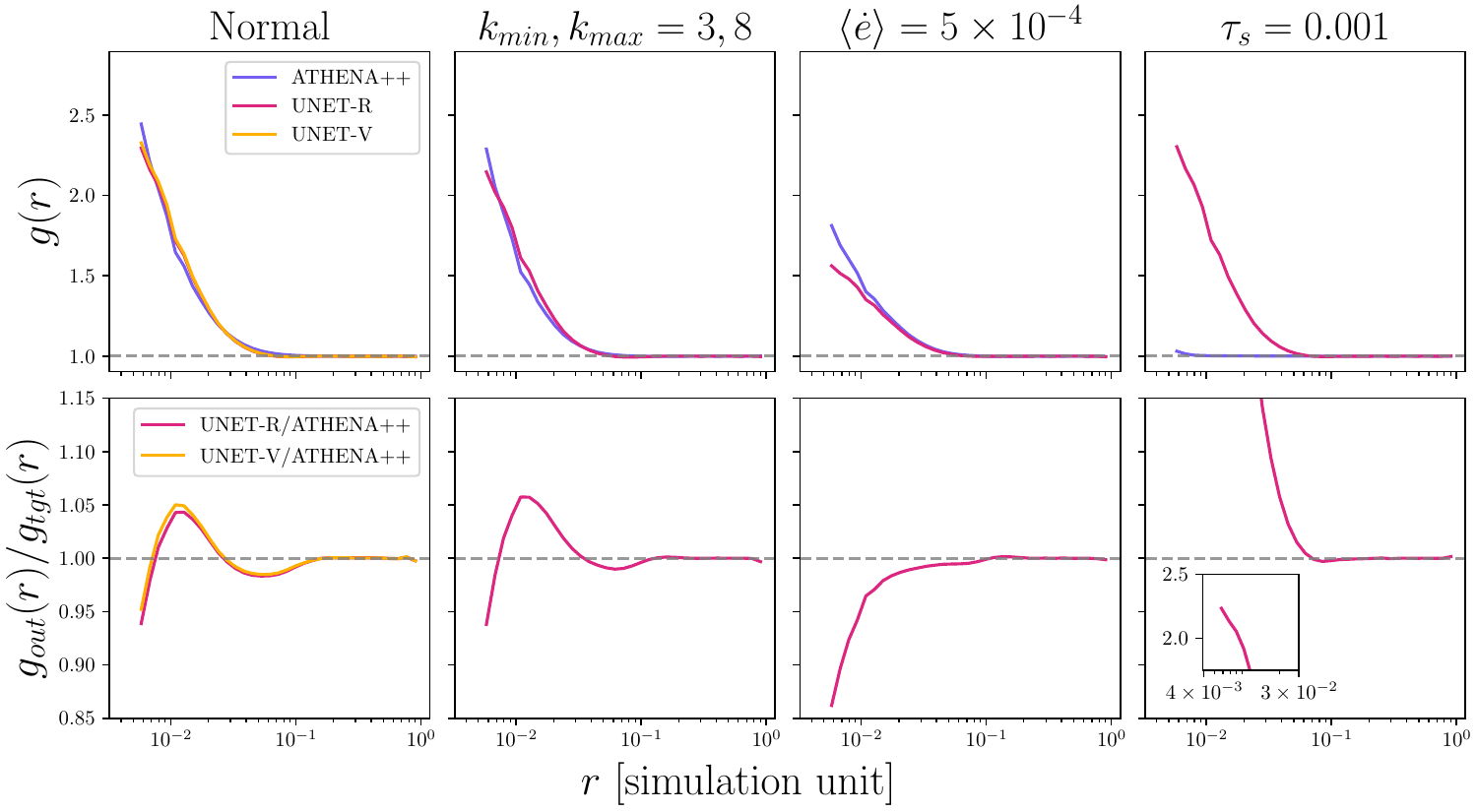}
    \caption{Predicted versus simulated strength of particle clustering as a function of spatial scale, quantified via the radial distribution function statistic (RDF). From left to right, the RDF plots show different simulation setups. The left-most panels show the fiducial setup, while the remaining panels show the results of generalization tests. (See \S\ref{sec:forced_turbulence_simulation},\ref{sec:lagrangian_dust_particles} and \S\ref{sec:generalization} for details.)  {ATHENA++} shows the RDF computed directly from the simulation, {UNET-R} is the U-Net output trained with relative velocity between gas and dust, and {UNET-V} is the U-Net output trained directly on dust velocity. The first row compares the RDF of the  simulation output and the U-Net prediction. The second row shows the ratio between the outputs and the target to better quantify the error. The zoom-in plot in the lower-right panel shows the peak of the error at small scales.}
    \label{fig:rdf_grouped}
\end{figure*}

In the left-most panel of Fig.~\ref{fig:rdf_grouped}, we show the RDFs of $\rho_p$ produced by the simulation (ATHENA++), and by the U-Nets (UNET-V, UNET-R). Recall, UNET-V is the U-Net trained directly on $v_p$, while UNET-R is trained on $v_{rel}$. The features of the RDFs reflect what we see in the slice plots. The simulation and U-Nets agree well for $r \gtrsim 0.01$, indicating that the large scale features are captured. For $r\approx 7\times10^{-3}$, a dip of roughly $5\%$ is observed, indicating the underestimation of $\rho_p$ at small scales. The over-prediction for $r\gtrsim 7\times 10^{-3}$ might be due to the smoothed density around the peaks. Overall, the error in RDF statistics is bounded by $\pm 5\%$. Although we have picked only one frame for discussion, the features observed are general. Discussion on the consistency across frames can be found in Appendix~\ref{sec:consistency_across_frames}. 

\section{Prediction of particle velocity field}
\label{sec:normal_particle_velc_prediction}

\subsection{Visual Comparison}
\begin{figure*}
    \centering
    \includegraphics[width=0.90\textwidth]{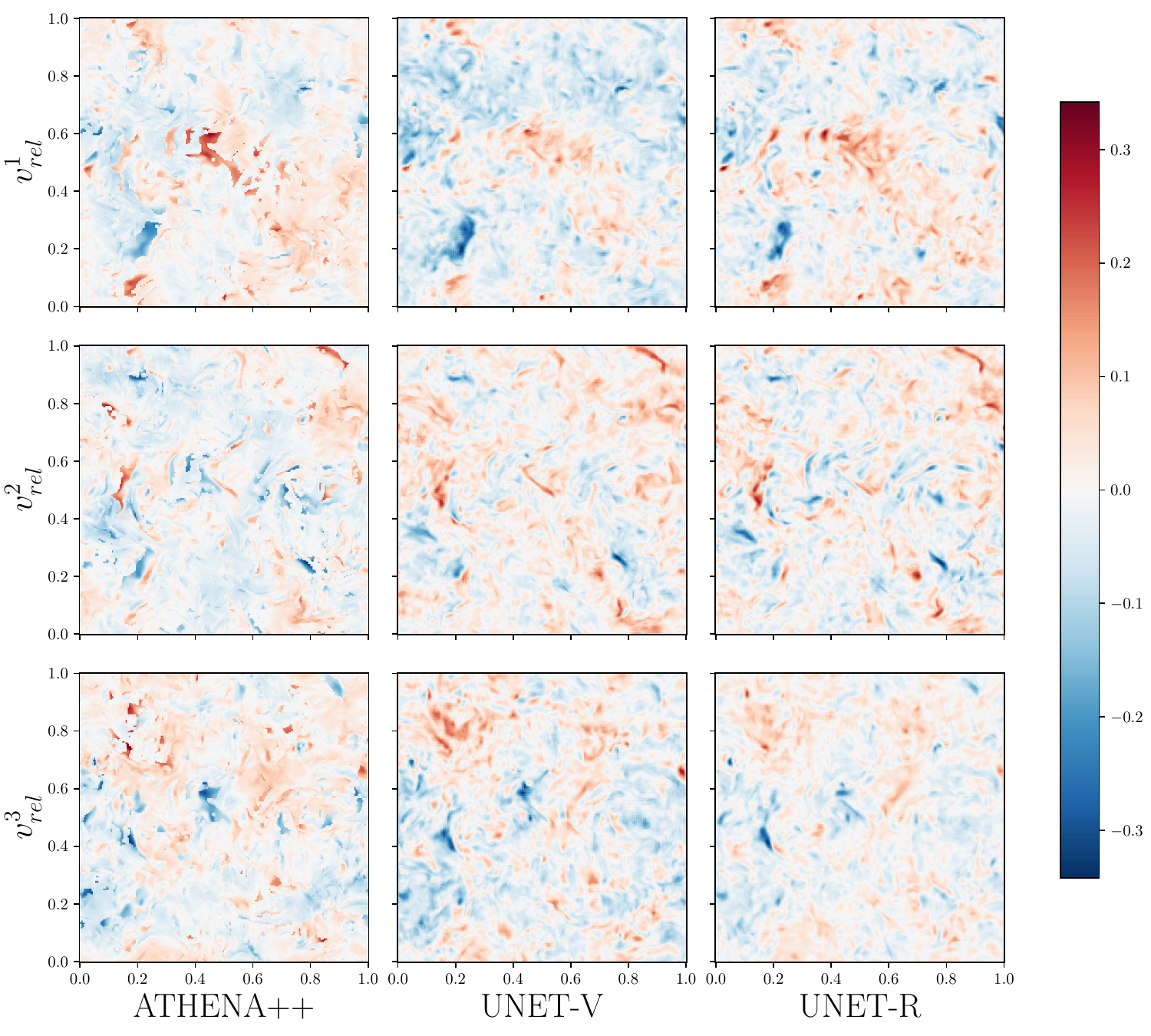}
    \caption{Prediction of the spatial distribution of particle relative velocity ($v_g-v_p$), in domains of isotropic forced turbulence. The figure compares the ground truth, obtained with {\sc ATHENA++}, with predictions from two deep learning models: UNET-V trained directly with particle velocity $v_p$, and UNET-R with relative velocity $v_{rel}$. The columns represent different setups (ATHENA++, UNET-V, UNET-R) and the rows show the 3 components of the velocity field. Each panel is a slice through the three-dimensional domain.}
    \label{fig:vrel_slice_comp}
\end{figure*}

In Fig.~\ref{fig:vrel_slice_comp}, we display slices of the three spatial components of the relative velocity field $v_{rel}$ produced by the simulation and by the U-Nets. Recall that the particle velocity $v_p$ can be thought of as having two contributions: (i) a gas velocity term $v_g$, which represents the velocity of the underlying flow driving the dust, and (ii) a relative velocity term $v_{rel} = v_p - v_g$, which measures the deviation of the particle velocity from the gas velocity due to the imperfect coupling between them. To prevent $v_g$ from overshadowing $v_{rel}$ in visual comparisons, we show $v_{rel}$.

From Fig.~\ref{fig:vrel_slice_comp}, we see that both UNET-V and UNET-R produce $v_{rel}$ fields that are qualitatively similar to that of the ATHENA++ ground truth. It is difficult to tell which is better solely by visual comparison. For example, at around $(v_{rel}^{3},0.5,0.6)$, UNET-V produce a sharper output for places where that component peaks, whereas at around $(v_{rel}^{3},0.1-0.3, 0.6-0.9)$, UNET-R captures the zeros of the field better. This motivates us to use more sophisticated statistics in \S\ref{sec:rel_velc_normal} and \S\ref{sec:rel_velc_rad_normal}, to assess which of them performs better.

\subsection{Relative velocity $\langle w^2 \rangle^{1/2}$}
\label{sec:rel_velc_normal}
The relative particle velocity $\langle w^2 \rangle^{1/2}$ is a commonly studied statistic of particle clustering in turbulence. Following \citet{ishihara2018dust}, we define it as, 
\begin{equation}
    \langle w^2 \rangle  = \frac{1}{N_p}\sum_{|\vec{r_1} - \vec{r_2}| = r} |\vec{V}_1 - \vec{V}_2|^2,
    \label{eq:rel_vel}
\end{equation}
where $r$ is the separation between the particle pairs, $\vec{V}_1$, $\vec{V}_2$ are the velocities of the particles located at $\vec{r}_1,\vec{r}_2$ and $N_p$ is the number of pairs of particles. 

To compute Eq.~\ref{eq:rel_vel} on the grid, we convert the average into a weighted one using the relevant particle density $\rho_p$ as weight. More precisely, let $p$ be a grid cell and $\text{C}_p(r_1,r_2)$ be the set of all grid cells whose center point lies inside the spherical shell with inner and outer radius $r_1$ and $r_2$ that is centered at $p$. Then a grid average of some quantity $Q$, weighted using particle density $\rho_p$, is defined as,
\begin{equation}
    \langle Q\rangle^{\rho_p}_\text{grid} = \frac{\sum_{q\in C_p(r,r+\Delta r) }\rho_p(q) Q(q)}{\sum_{q\in C_p(r,r+\Delta r)} \rho_p(q)},
\end{equation}
where $\rho_p(q)$ and $Q(q)$ are understood as the grid values at cell $q$. Using this definition, we write $\langle w^2_\text{tgt}\rangle^{1/2} = \left[\langle w^2_\text{tgt}\rangle^{\rho^\text{tgt}}_\text{grid}\right]^{1/2}$ for the ground truth relative velocity and $\langle w^2\rangle^{1/2}=\langle w^2_\text{out}\rangle^{1/2} = \left[\langle w^2_\text{out}\rangle^{\rho^\text{out}}_\text{grid}\right]^{1/2}$ for the output velocity field. 

Unlike grid RDFs, computing these quantities on the grid is relatively expensive. To compute the statistic  for the whole grid $r\in [256^{-1},1]$, we would need to check every grid point against every other grid point. If $N$ is the number of grid points along one axis, this would imply a complexity $\approx (N^3)^2=N^6$. %With $N=256$, we need to do $\approx 2.8\times 10^{14}$ checks, whereas the FFT accelerated RDF code only takes $\approx N^3 \log(N^3) \approx 4.02\times 10^{8}$ operations.
To reduce the computational cost, we restrict $r\in[256^{-1},10^{-1}]$ for all relative velocity statistics. This is not a serious restriction because for many applications it is the accuracy of relative velocities at small scales that is the most important. Furthermore, from the estimated clustering scale in \S\ref{sec:lagrangian_dust_particles}, we see that the upper bound of the radii $r=10^{-1} \gg \ell_\eta \sim 10^{-3}$, so the relative velocity curve for $r>10^{-1}$ is likely not relevant for our application. Again, possible errors introduced by grid averaging are explored in Appendix~\ref{sec:grid_stats_vs_part_stats}.

\begin{figure*}
    \centering
    \includegraphics[width=\textwidth]{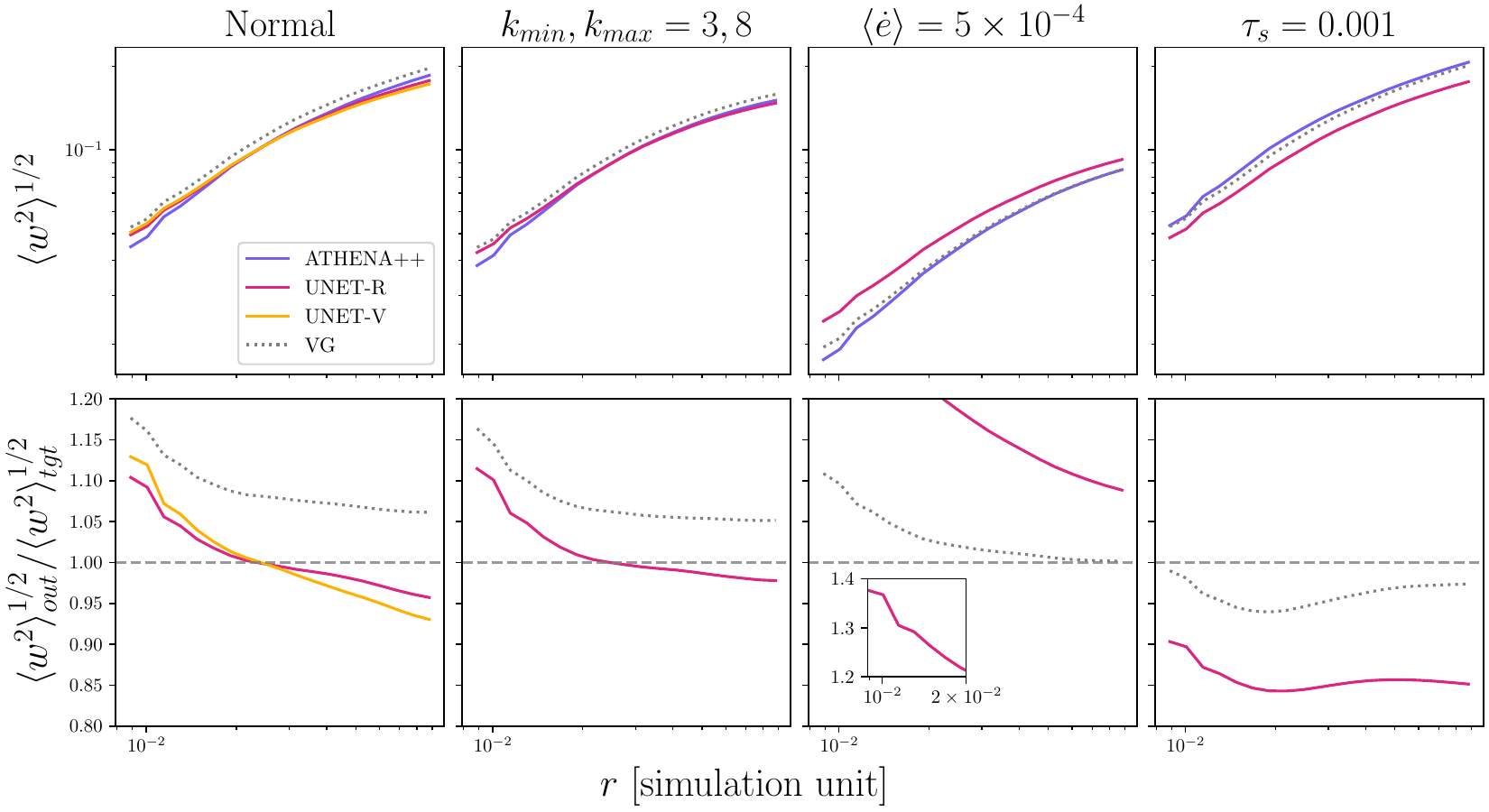}
    \caption{Comparison of relative velocity between dust particles $\langle w^2 \rangle^{1/2}$ for different simulation setups (see \S\ref{sec:forced_turbulence_simulation},\ref{sec:lagrangian_dust_particles} and \S\ref{sec:generalization} for details). {ATHENA++} represents the relative velocity computed from the simulation, {UNET-R} is the U-Net output trained with relative dust velocity, {UNET-V} is the U-Net output trained directly on dust velocity, and {VG} is the average gas velocity weighted by particle density (i.e. the relative velocity curve one would have obtained if the particles follow perfectly the flow). The left-most panels show the fiducial results, the other panels show the results of generalization tests.}
    \label{fig:vp_rel_grouped}
\end{figure*}

We plot $\langle w^2\rangle^{1/2}$ and its error as a function of $r$ in Fig.~\ref{fig:vp_rel_grouped}. The upper left panel shows that the relative particle velocity fields produced by the UNET-V, UNET-R, and the ground truth (ATHENA++) are similar in shape, and the lower left panel shows that both models have their error bounded by approximately $15\%$ above and $7\%$ below. For $r \gtrsim 2.5 \times 10^{-2}$, both models under-predict the relative velocity, while for $r\lesssim 2.5 \times 10^{-2}$ both over-predict. The absolute error for the UNET-R model is always less than that of UNET-V at all length scales. In particular, the errors in UNET-R's prediction are $\approx 2.5\%$ less than UNET-V's at the smallest and largest scales. Furthermore, the region around the point where UNET-R's error curve crosses unity, especially that toward the tail of the curve, is much flatter than that of UNET-V's. This indicates the error varies less with separation in UNET-R.

The results above support our hypothesis that targeting the relative velocity aids in the network training. UNET-R can capture the radial dependence of relative velocity in the predicted particle velocity field within a maximum error of approximately 10\%. Although this error is systematic and scale-dependent rather than random, we consider this to be good performance. 

\subsection{Relative radial velocity $\langle w^2_\text{r} \rangle^{1/2}$}
\label{sec:rel_velc_rad_normal}
\begin{figure*}
    \centering
    \includegraphics[width=\textwidth]{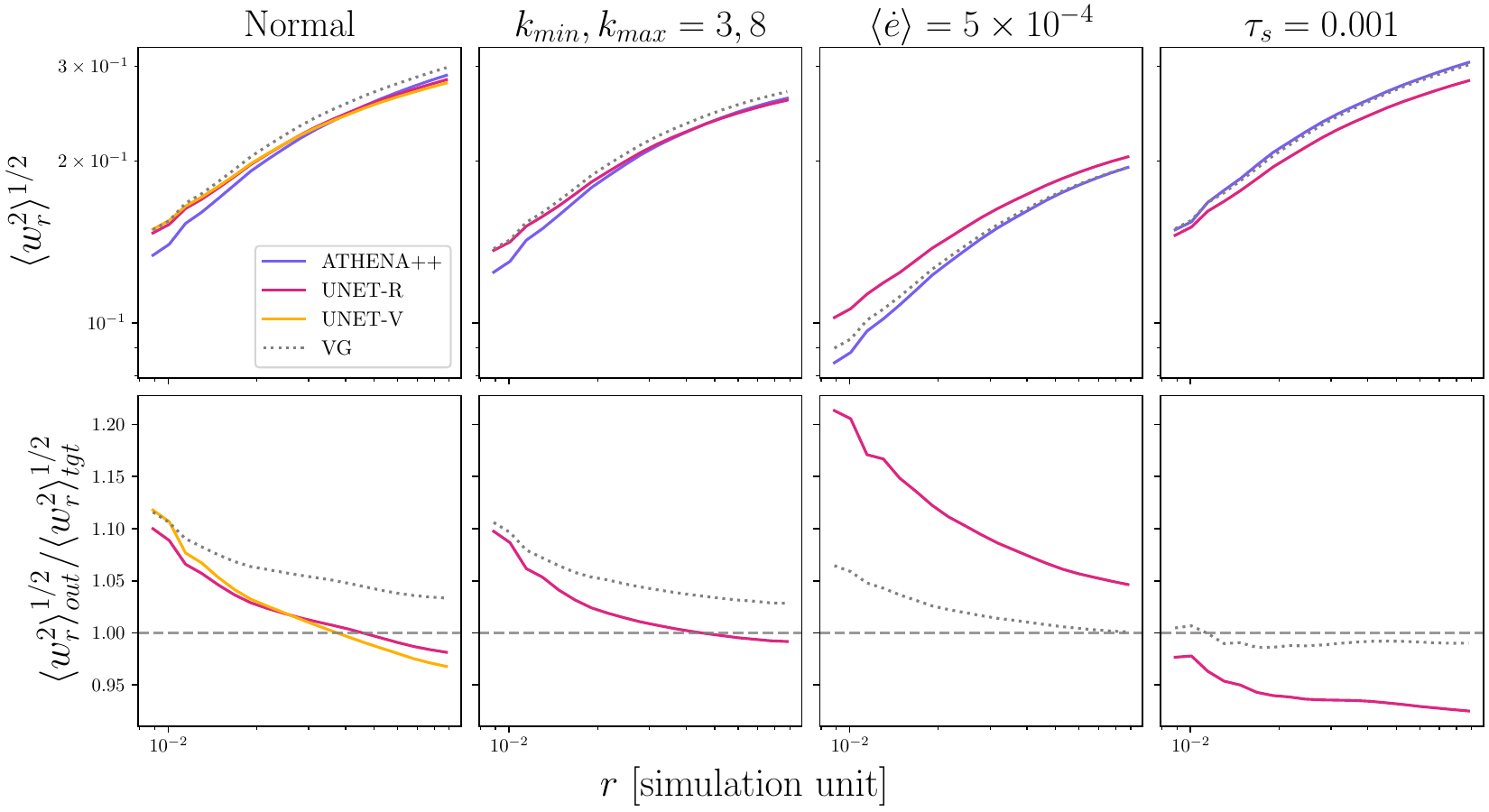}
    \caption{Comparison of relative radial velocity $\langle w^2_r \rangle^{1/2}$ for different simulation setups (see \S\ref{sec:forced_turbulence_simulation},\ref{sec:lagrangian_dust_particles} and \S\ref{sec:generalization} for details). Refer to Fig.~\ref{fig:vp_rel_grouped} for the description on the abbreviated labels.}
    \label{fig:vp_rel_rad_grouped}
\end{figure*}

Similarly, the root-mean-square relative radial velocity $\langle w_\text{r}^2\rangle^{1/2}$ is defined as,
\begin{equation}
    \langle w_\text{r}^2 \rangle  = \frac{1}{N_p}\sum_{|\vec{r_1} - \vec{r_2}| = r} \left(\left(\vec{V}_1 - \vec{V}_2\right)\cdot \hat{\eta}\right)^2,
\end{equation}
where all the symbols retain the same meaning as in Eq.~\ref{eq:rel_vel}, and $\hat{\eta} = (\vec{r}_1 - \vec{r}_2)/|\vec{r}_1 - \vec{r}_2|$. Using the same weighting procedure as in \S\ref{sec:rel_velc_normal}, we again computed $\langle w^2_\text{r}\rangle$ on ATHENA++, UNET-V and UNET-R data. The results are shown in Fig.~\ref{fig:vp_rel_rad_grouped}.

In the figure, we see that both the relative particle velocity $\langle w_{r}^2\rangle^{1/2}$ predicted by UNET-R and UNET-V have similar shapes to the target. However, the performance of the U-Nets varies at different scales. For $r\gtrsim 2.5\times 10^{-2}$ both models perform similarly and predict a radial velocity around $\approx 2.5\%$ above and $\approx 5\%$ below the target respectively. The two models over-predict at scales $r\lesssim 4\times 10^{-2}$, being $\approx 10\%$ and $\approx 12.5$ larger respectively. Unlike $\langle w^2\rangle^{1/2}$, the error curves of the models do not intersect at unity. However, UNET-R's error remains  smaller than UNET-V's across most scales, except for a small region around $r\approx 3\times 10^{-2}$.

Also, while both UNET-R and UNET-V overpredict the radial velocity at small scales, that of UNET-V is much closer to the curve labeled VG---the average gas velocity weighted by particle density---than UNET-R. This indicates that UNET-V is essentially outputting the gas velocity $v_g$ as the particle velocity at small scales, instead of learning the particles' deviation from the gas flow. This further demonstrates that training on $v_{rel}$ is effective in helping the network to single out how the particle velocity differs from the background flow. 

Finally, unlike the relative velocity, which shows a flat part at intermediate scales, there are no obvious flat regions on UNET-V's error curve. This suggests the network in general predicts the direction of $v_g$ less accurately than its amplitude. 

\subsection{Error estimation}
The predicted particle density and velocity fields, and their associated statistics, show good but not perfect agreement with the ground truth derived from the simulation data. The errors are systematic and scale-dependent. It is of interest to ask: how much of this error is due to uncertainty in the deep learning model? Addressing this question is difficult, and various approximate methods have been used in the machine learning community. We have experimented with two popular frameworks, (i) Monte-Carlo Dropout \citep{gal2016dropout,zhu2017deep} and (ii) stochastic weight averaging–Gaussian (SWAG) \citep{maddox2019simple}. Details are given in Appendix~\ref{sec_errors}. In summary, we find evidence using the SWAG approach that model uncertainty has a larger effect on the velocity predictions than on the density predictions. However, the magnitude of the estimated model uncertainty is significantly smaller than the size of the discrepancy between the model predictions and the ground truth. The source of the discrepancy thus remains unclear, but it is clearly of a systematic nature. 

\section{Generalization Tests}
\label{sec:generalization}

%Ideally, we would like our network, trained only with a limited set of samples, to generalize its' prediction over different energy injection rates $e_\text{inj}$, turbulent driving spectrum $(k_\text{low},k_\text{high})$, and turbulence length scales $\ell_\eta$. However, since the train set we use above covers only one particular combinations these physically important parameters, it is not likely it will generalize well. 

Our deep learning model, trained on a single simulation of forced turbulence with a specified set of parameters, cannot be expected to generalize to distinctly different regimes of particle-gas coupling. However, it {\em may} have learned enough about how particles interact with turbulent fluids to behave robustly when confronted by modestly different turbulent conditions. To test our model's ability to generalize, we have designed three test cases with physical parameters different from the fiducial case discussed so far. The test cases are: (1) turbulence driven with a different range of cut off wavenumbers $k_{min},k_{max}$; (2) turbulence driven using a different energy injection rate of $\langle\dot{e}\rangle = 5\times 10^{-4}$; (3) particles with stopping time $\tau_s$ 100 times smaller than the original simulation. ATHENA++ simulations with these different parameters, but otherwise similar to the fiducial run, were performed for each test case.

For each of the test cases, we feed their simulation frames into the UNET-R trained from the fiducial run without any additional training, and produce field predictions for $\rho_p$ and $v_p$. Then, using the same statistics code in \S\ref{sec:normal_stat_rdf}, \S\ref{sec:rel_velc_normal} and \S\ref{sec:rel_velc_rad_normal}, we compute the relevant statistics from the field outputs and compare them with the ground truth. 

\subsection{Different wavenumber cut-off}
\label{sec:generalize_k_cutoff}
\begin{figure}
    \centering
    \includegraphics[width=\linewidth]{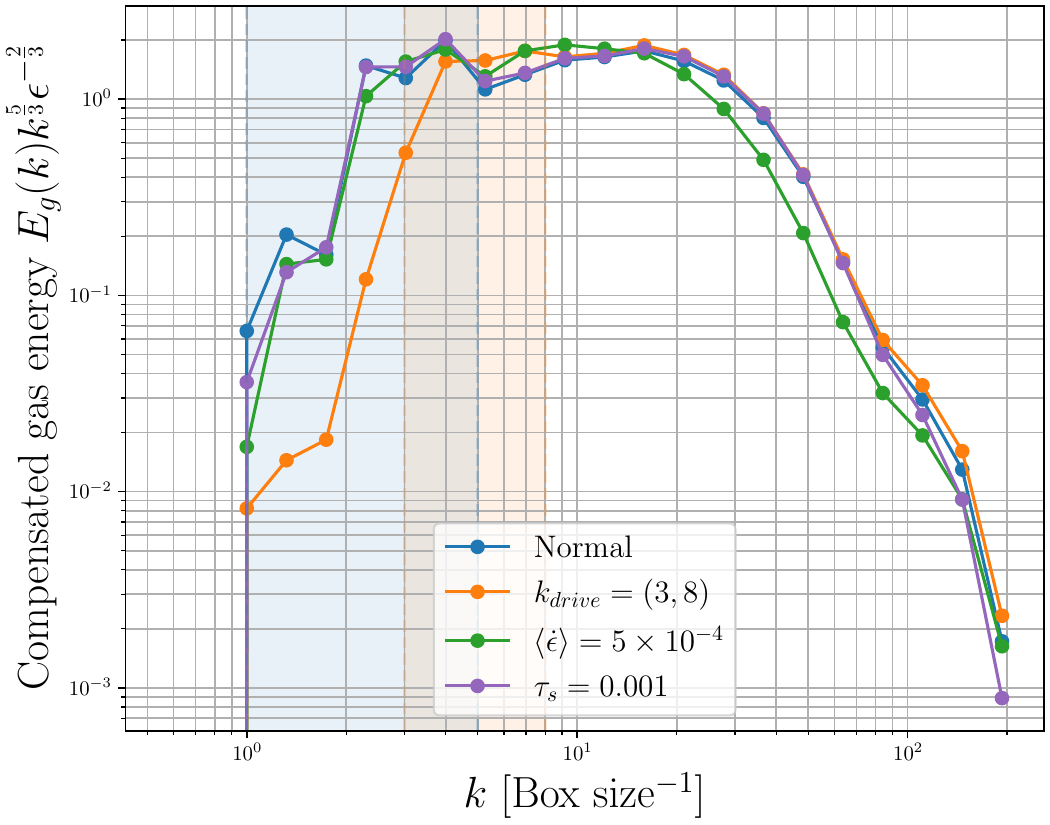}
    \caption{\revTwoEdit{Compensated energy spectrum of driving flow $v_g$ for different simulation setups. The color bands illustrate the range of $k$ used in driving the flow for each case and the line shows the resulting power spectrum of the gas kinetic energy.}}
    \label{fig:vg_power_spectrum}
\end{figure}

In this generalization set, different minimum and maximum cutoff of driving wavenumbers $k$ are used. Instead of the original $(k_{min},k_{max})=(1,5)$ combination, we remove the largest scale driving and use the combination $(3,8)$. The resulting gas velocity power spectrum is illustrated in Fig.~\ref{fig:vg_power_spectrum}. The dip in power at around $k=1-3$ confirms that the new flow has less power in large scales compared to the original flow.

The RDF is computed and plotted in the second column in Fig.~\ref{fig:rdf_grouped}.In the figure, we notice that the U-Net predicted RDF resembles closely that of ATHENA++. Furthermore, from the ratio plot, we see the discrepancy between the two densities takes a similar shape as that produced on the original validation set. 

As mentioned in \S\ref{sec:phy_of_clustering}, the particle clustering is most responsive to eddies with size $\ell \approx \text{St}^{3/2} \ell_\eta$. Therefore, the absence  of large-scale driving should not affect the resulting clustering. This test indicates the network can pick up the correct scales in the power spectrum responsible for the clustering.

The relative velocity $\langle w^2\rangle^{1/2}$ and the radial relative velocity $\langle w_r^2\rangle^{1/2}$ behave similarly to the density. Due to the energy cascade inherent to Kolmogorov turbulence, it is expected that the removal of large-scale driving in this test case should not affect the small-scale relative velocity. Therefore a similar target relative velocity profile is expected. This is verified in the second plot in Fig.~\ref{fig:vp_rel_grouped} and Fig.~\ref{fig:vp_rel_rad_grouped}. Furthermore, the network also outputs a similar relative velocity as the target. The relative velocity of the predicted $v_p$ is relatively flat for $r\gtrsim 2\times 10^{-2}$ and raises for scales smaller than $r\lesssim 2\times 10^{-2}$. The ``flat" part is at most $\approx 3\%$ lower than the ground truth and the ``peak" is at most $\approx 12\%$ above the target just like that in the normal validation set. The predicted relative velocity also differ from VG by $\approx 5\%$, which indicates the network can discriminate gas velocities from dust velocities. This shows that the network correctly ignores the large-scale driving that is irrelevant to the relative velocity on small-scales and identifies the wave-numbers that are relevant for the problem. As a result the network is able to predict the relative velocity field with similar quality as the one it is trained on.

The radial component shows very similar character to the relative velocity. From Fig.~\ref{fig:vp_rel_rad_grouped}, we see that at the largest scale the prediction is $\approx 1\%$ below the ground truth and at the smallest scale it is $\approx 10\%$ larger. However, unlike the $\langle w^2\rangle^{1/2}$ predictions, the difference between the predicted radial velocity and VG is smaller in this case. Still, the predictions made by the network have similar quality as that in the fiducial case. This shows that for all statistics above, the network is indeed able to find the correct wave-numbers that are responsible for the relative velocity.

\subsection{Different energy injection rate}
\label{sec:generalize_dedt}
In this generalization test, we drive the turbulence with an energy injection rate that is 10 times smaller than that of the fiducial case ($\langle \epsilon \rangle =5\times 10^{-3}$). As shown in \S\ref{sec:phy_of_clustering}, the length scale at which the strongest clustering occurs is given as $\ell_\tau \approx \langle \epsilon \rangle^{1/2} \tau_s^{3/2}$. With $\langle \epsilon \rangle  \to 0.1\langle \epsilon \rangle $, the clustering length scale should be $0.1^{1/2} \approx 0.32$ times the original. Hence, the dust clustering structure becomes finer and the RDF should rise at smaller scales. Since our statistics are limited to grid resolution $r\approx 256^{-1}$, the RDF instead appears flatter. This is reflected in the RDF of the simulated ground truth in the third subplot of Fig.~\ref{fig:rdf_grouped}, where the peak $g(r)$ value is reduced from $2.25$ to $1.8$. \revTwoEdit{The lower energy injection rates also pushes the dissipation scale to lower $k$. This is demonstrated qualitatively in Fig.~\ref{fig:vg_power_spectrum}, where the setup's spectral energy starts falling off at a lower $k$ compared to the other setups.}

However, the network output does not show similar scaling. As shown in the figure, starting from $r\lesssim 10^{-1}$, the network consistently underpredicts the RDF. From $10^{-2}\lesssim r\lesssim 10^{-1}$, the prediction is $\lesssim 3\%$ from the target. However, the deviation quickly grows for scales $r\lesssim 10^{-2}$, reaching an error of $\approx 15\%$ at a length of $\approx 5\times 10^{-3}$. This shows that while the network is able to partially learn the dependency of clustering on $\langle \epsilon \rangle $, it is unable to account for the enhanced clustering at smaller scales. We note that all predicted RDFs go to 1 as $r\gtrsim 10^{-1}$. This shows that the predicted dust density indeed goes to the average density at large scale, as expected from the definition of the RDF.

Similarly, reducing the energy injection rate will affect the gas velocity scale. By Kolmogorov scaling, the gas velocity in the inertial range should scale as $v_g \propto \langle \epsilon \rangle^{1/3}$. Since the dust is driven by the gas via aerodynamic drag, we expect the relative velocity of the particles to scale partly as $\propto \langle \epsilon \rangle^{1/3}$ too. Therefore, reducing $\langle \epsilon \rangle  \to 0.1\langle \epsilon \rangle $ should scale down the velocity by a factor of $10^{1/3} \approx 2.15$. This scaling is roughly reflected in the third columns of Figs.~\ref{fig:vp_rel_grouped},\ref{fig:vp_rel_rad_grouped}. The minimum and maximum target relative velocity scale of the fiducial average relative velocity and radial velocity are $\approx 5\times 10^{-2} - 2\times 10^{-1}$ and $\approx 1.3\times 10^{-1} - 3\times10^{-1}$ respectively, while that in this generalization test is $\approx 2\times 10^{-2} - 8\times 10^{-2}$ and $\approx 9\times 10^{-2} - 2 \times 10^{-1}$ respectively. Therefore, the target relative velocities are all smaller than the fiducial case by approximately the expected factor, and for small scale the difference roughly corresponds to the scaling we derived above. However, the network does not display similar scaling. As shown in Fig.~\ref{fig:vp_rel_grouped}, the network always over-predicts the relative velocity by a factor of at least $\approx 10\%$. Furthermore, the deviation grows steadily as the scale gets smaller, peaking at $\approx 40\%$ error at the smallest scale. 

Analogous results is shown in radial velocity statistics in Fig.~\ref{fig:vp_rel_rad_grouped}. At all scales, the network over-predicts the statistics by at least $6\%$. This deviation continues to grow for smaller $r$ and an $\approx 20\%$ overshoot in radial velocity at grid-scale is observed. Furthermore, comparing the VG line and the network output in Fig.~\ref{fig:vp_rel_grouped} and Fig.~\ref{fig:vp_rel_rad_grouped}, we see that the predicted particle velocities are much higher than what would have been obtained if all dust was perfectly following the flow (VG). Hence, the network is likely predicting particle relative velocities that correspond to simulations with a higher velocity scale. This reflects the network's inability to extract correct injection energy scales from the gas velocity field. While this is a failure of the model, it is expected because the network is trained only with one energy injection rate.

\subsection{Different particle stopping time $\tau_s$}
\label{sec:generalize_diff_stop_time}
Finally, we briefly discuss the results of a more challenging generalization test, in which we changed the dust stopping time $\tau_s$ for the particles to 0.001, a hundred times smaller than that of the original simulation. \revTwoEdit{On the other hand, the gas spectrum remains largely unchanged (See Fig.~\ref{fig:vg_power_spectrum})}. Since $\ell_\tau \approx \tau_s^{3/2}\epsilon^{1/2}$, this means the particle clustering scale will be scaled down to around $0.1 \%$ of the original flow. Again, by the same reasoning as \S\ref{sec:generalize_dedt}, a flat ground-truth RDF is expected. This is reflected in the rightmost plot in Fig.~\ref{fig:rdf_grouped}. The violet line, which represents the RDF of the target flow, is nearly flat and only deviates slightly from $1$ at $r\lesssim 10^{-2}$. The strong coupling means that the velocity of the particles in the simulation follows that of the gas closely at all scales.

In contrast to the less challenging generalization tests presented above, the large change in stopping time leads to a catastrophic failure in the prediction of the particle RDF. The results for the different velocity statistics, while not wrong by a similarly large factor, are systematically lower than the ground truth at all scales (See Fig.~\ref{fig:vp_rel_grouped} \& \ref{fig:vp_rel_rad_grouped}). These failures are to be expected because $\tau_s$ is a physically important parameter that our model has no knowledge of. They illustrate the importance of including simulations with different stopping times as part of the training set for future models.

\section{Discussion} 
\label{sec:conclusion}
In this work, we have quantified the ability of the U-Net machine learning architecture to predict the clustering of particles that are aerodynamically coupled to turbulent gas. Training data was generated using a recently developed particle module for the {\sc ATHENA++} code, with the included physics and parameters chosen to be appropriate for the problem of small-scale turbulent clustering of dust in protoplanetary disks. After training, using a loss function that targets the three-dimensional particle density and velocity fields, we find that the U-Net model yields satisfactory performance for gridded representations of the {\em statistical} metrics that are of greatest interest for planet formation---the Radial Distribution Function and particle relative velocity. Errors on these quantities are scale-dependent, but typically at the percent level. Smaller maximum errors are found for density rather than velocity statistics. The model also generates, of course, a fast approximate mapping from the fluid to the particle fields in three dimensions. This is a novel capability that is not, to our knowledge, possible using analytic methods.

The use of a U-Net architecture requires a gridded representation of the particle density, and the definition of a gridded (single-valued) velocity field. These characteristics lead directly to some of the limitations observed in our study. The Lagrangian representation of particles in the training simulations yields useful information on particle statistics on scales {\em smaller} than those of the fluid grid, which is discarded as a pre-processing step in our approach. For applications where the sole interest is the statistical properties of the particles, better results extending to smaller scales would likely be obtained by using a convolutional neural network to target directly the radial distribution and relative velocity functions.

The model fails to generalize to different friction times and energy injection rates, as expected given that these are key physical parameters and the network was only exposed to a single value of them. There are no obvious obstacles to training the network on a suite of simulations and thereby learning the dependence of particle clustering on physical input parameters \citep[see, e.g. analogous work in cosmology;][]{vn21}. The combination of three-dimensional spatial prediction of particle positions, together with the prediction of the particles' velocity relative to the gas, would then allow the calculation of particle collision rates and outcomes for arbitrary size distributions. Different approaches would be needed to incorporate the back-reaction of particles on the gas, an aspect ignored in our work but which is central to models of planetesimal formation \citep{youdin05,simon16}. Learned simulators for turbulent flows show considerable promise \citep{dang22}, and it would be interesting to apply similar methods to multi-phase fluid turbulence. The existence of new instabilities that rely on the two-way coupling between gas and particles \citep{youdin05,squire18} suggests that multi-phase turbulence would be a stringent test of such methods.

The work presented here explores arguably the simplest problem in two-fluid turbulence that has relevance to planet formation. Related problems are present on multiple other scales in protoplanetary disks and in other domains, including atmospheric science and experimental studies of turbulence. Our results suggest that deep learning offers a complementary approach to direct numerical simulations that can predict particle clustering at a lower computational cost.

\acknowledgments
We thank Jim Stone for helpful discussions during the development of this project. PJA acknowledges support from NASA TCAN award 80NSSC19K0639, and from award 644616 from the Simons Foundation. CCY and ZZ acknowledge support from NASA via the Emerging Worlds program (award 80NSSC20K0347) and the Astrophysics Theory Program (award 80NSSC21K0141). CCY acknowledges support from NASA TCAN award 80NSSC21K0497.

\bibliographystyle{aasjournal}
\bibliography{dustML}

\begin{thebibliography}{}
\expandafter\ifx\csname natexlab\endcsname\relax\def\natexlab#1{#1}\fi
\providecommand{\url}[1]{\href{#1}{#1}}
\providecommand{\dodoi}[1]{doi:~\href{http://doi.org/#1}{\nolinkurl{#1}}}
\providecommand{\doeprint}[1]{\href{http://ascl.net/#1}{\nolinkurl{http://ascl.net/#1}}}
\providecommand{\doarXiv}[1]{\href{https://arxiv.org/abs/#1}{\nolinkurl{https://arxiv.org/abs/#1}}}

\bibitem[{{ALMA Partnership} {et~al.}(2015){ALMA Partnership}, {Brogan},
  {P{\'e}rez}, {Hunter}, {Dent}, {Hales}, {Hills}, {Corder}, {Fomalont},
  {Vlahakis}, {Asaki}, {Barkats}, {Hirota}, {Hodge}, {Impellizzeri}, {Kneissl},
  {Liuzzo}, {Lucas}, {Marcelino}, {Matsushita}, {Nakanishi}, {Phillips},
  {Richards}, {Toledo}, {Aladro}, {Broguiere}, {Cortes}, {Cortes}, {Espada},
  {Galarza}, {Garcia-Appadoo}, {Guzman-Ramirez}, {Humphreys}, {Jung}, {Kameno},
  {Laing}, {Leon}, {Marconi}, {Mignano}, {Nikolic}, {Nyman}, {Radiszcz},
  {Remijan}, {Rod{\'o}n}, {Sawada}, {Takahashi}, {Tilanus}, {Vila Vilaro},
  {Watson}, {Wiklind}, {Akiyama}, {Chapillon}, {de Gregorio-Monsalvo}, {Di
  Francesco}, {Gueth}, {Kawamura}, {Lee}, {Nguyen Luong}, {Mangum}, {Pietu},
  {Sanhueza}, {Saigo}, {Takakuwa}, {Ubach}, {van Kempen}, {Wootten},
  {Castro-Carrizo}, {Francke}, {Gallardo}, {Garcia}, {Gonzalez}, {Hill},
  {Kaminski}, {Kurono}, {Liu}, {Lopez}, {Morales}, {Plarre}, {Schieven},
  {Testi}, {Videla}, {Villard}, {Andreani}, {Hibbard}, \& {Tatematsu}}]{alma14}
{ALMA Partnership}, {Brogan}, C.~L., {P{\'e}rez}, L.~M., {et~al.} 2015, \apjl,
  808, L3, \dodoi{10.1088/2041-8205/808/1/L3}

\bibitem[{{Andrews}(2020)}]{andrews20}
{Andrews}, S.~M. 2020, \araa, 58, 483,
  \dodoi{10.1146/annurev-astro-031220-010302}

\bibitem[{{Andrews} {et~al.}(2018){Andrews}, {Huang}, {P{\'e}rez}, {Isella},
  {Dullemond}, {Kurtovic}, {Guzm{\'a}n}, {Carpenter}, {Wilner}, {Zhang}, {Zhu},
  {Birnstiel}, {Bai}, {Benisty}, {Hughes}, {{\"O}berg}, \& {Ricci}}]{andrews18}
{Andrews}, S.~M., {Huang}, J., {P{\'e}rez}, L.~M., {et~al.} 2018, \apjl, 869,
  L41, \dodoi{10.3847/2041-8213/aaf741}

\bibitem[{{Barge} \& {Sommeria}(1995)}]{barge95}
{Barge}, P., \& {Sommeria}, J. 1995, \aap, 295, L1.
\newblock \doarXiv{astro-ph/9501050}

\bibitem[{{Bhatnagar} {et~al.}(2018){Bhatnagar}, {Gustavsson}, {Mehlig}, \&
  {Mitra}}]{bhatnagar18}
{Bhatnagar}, A., {Gustavsson}, K., {Mehlig}, B., \& {Mitra}, D. 2018, \pre, 98,
  063107, \dodoi{10.1103/PhysRevE.98.063107}

\bibitem[{{Blum} \& {Wurm}(2008)}]{blum08}
{Blum}, J., \& {Wurm}, G. 2008, \araa, 46, 21,
  \dodoi{10.1146/annurev.astro.46.060407.145152}

\bibitem[{Bracco {et~al.}(1999)Bracco, Chavanis, Provenzale, \&
  Spiegel}]{bracco1999particle}
Bracco, A., Chavanis, P., Provenzale, A., \& Spiegel, E. 1999, Physics of
  Fluids, 11, 2280

\bibitem[{{Chambers}(2010)}]{chambers10}
{Chambers}, J.~E. 2010, \icarus, 208, 505, \dodoi{10.1016/j.icarus.2010.03.004}

\bibitem[{Ching(2014)}]{ching2014statistics}
Ching, E.~S. 2014, Statistics and scaling in turbulent Rayleigh-Benard
  convection (Springer)

\bibitem[{{Dang} {et~al.}(2022){Dang}, {Hu}, {Cranmer}, {Eickenberg}, \&
  {Ho}}]{dang22}
{Dang}, Y., {Hu}, Z., {Cranmer}, M., {Eickenberg}, M., \& {Ho}, S. 2022, arXiv
  e-prints, arXiv:2207.04616.
\newblock \doarXiv{2207.04616}

\bibitem[{Davydzenka \& Tahmasebi(2022)}]{davydzenka_tahmasebi_2022}
Davydzenka, T., \& Tahmasebi, P. 2022, Journal of Fluid Mechanics, 938, A20,
  \dodoi{10.1017/jfm.2022.174}

\bibitem[{Gal \& Ghahramani(2016)}]{gal2016dropout}
Gal, Y., \& Ghahramani, Z. 2016, in international conference on machine
  learning, PMLR, 1050--1059

\bibitem[{{Hartlep} \& {Cuzzi}(2020)}]{hartlep20}
{Hartlep}, T., \& {Cuzzi}, J.~N. 2020, \apj, 892, 120,
  \dodoi{10.3847/1538-4357/ab76c3}

\bibitem[{Huang \& Bai(2022)}]{huang2022multifluid}
Huang, P., \& Bai, X.-N. 2022, The Astrophysical Journal Supplement Series,
  262, 11

\bibitem[{{Ishihara} {et~al.}(2018){Ishihara}, {Kobayashi}, {Enohata},
  {Umemura}, \& {Shiraishi}}]{ishihara18}
{Ishihara}, T., {Kobayashi}, N., {Enohata}, K., {Umemura}, M., \& {Shiraishi},
  K. 2018, \apj, 854, 81, \dodoi{10.3847/1538-4357/aaa976}

\bibitem[{Ishihara {et~al.}(2018)Ishihara, Kobayashi, Enohata, Umemura, \&
  Shiraishi}]{ishihara2018dust}
Ishihara, T., Kobayashi, N., Enohata, K., Umemura, M., \& Shiraishi, K. 2018,
  The Astrophysical Journal, 854, 81

\bibitem[{{Johansen} {et~al.}(2014){Johansen}, {Blum}, {Tanaka}, {Ormel},
  {Bizzarro}, \& {Rickman}}]{johansen14}
{Johansen}, A., {Blum}, J., {Tanaka}, H., {et~al.} 2014, in Protostars and
  Planets VI, ed. H.~{Beuther}, R.~S. {Klessen}, C.~P. {Dullemond}, \&
  T.~{Henning}, 547, \dodoi{10.2458/azu\_uapress\_9780816531240-ch024}

\bibitem[{Johansen {et~al.}(2015)Johansen, Low, Lacerda, \&
  Bizzarro}]{johansen15}
Johansen, A., Low, M.-M.~M., Lacerda, P., \& Bizzarro, M. 2015, Science
  Advances, 1, e1500109

\bibitem[{{Johansen} {et~al.}(2009){Johansen}, {Youdin}, \&
  {Klahr}}]{johansen09}
{Johansen}, A., {Youdin}, A., \& {Klahr}, H. 2009, \apj, 697, 1269,
  \dodoi{10.1088/0004-637X/697/2/1269}

\bibitem[{Kendall \& Gal(2017)}]{kendall2017uncertainties}
Kendall, A., \& Gal, Y. 2017, Advances in neural information processing
  systems, 30

\bibitem[{Kingma \& Ba(2014)}]{kingma2014adam}
Kingma, D.~P., \& Ba, J. 2014, arXiv preprint arXiv:1412.6980

\bibitem[{{Kolmogorov}(1941)}]{kolmogorov41}
{Kolmogorov}, A. 1941, Akademiia Nauk SSSR Doklady, 30, 301

\bibitem[{Kolmogorov(1991)}]{kolmogorov1991local}
Kolmogorov, A.~N. 1991, Proceedings of the Royal Society of London. Series A:
  Mathematical and Physical Sciences, 434, 9

\bibitem[{{Krapp} {et~al.}(2019){Krapp}, {Ben{\'\i}tez-Llambay}, {Gressel}, \&
  {Pessah}}]{krapp19}
{Krapp}, L., {Ben{\'\i}tez-Llambay}, P., {Gressel}, O., \& {Pessah}, M.~E.
  2019, \apjl, 878, L30, \dodoi{10.3847/2041-8213/ab2596}

\bibitem[{{Krapp} {et~al.}(2020){Krapp}, {Youdin}, {Kratter}, \&
  {Ben{\'\i}tez-Llambay}}]{krapp20}
{Krapp}, L., {Youdin}, A.~N., {Kratter}, K.~M., \& {Ben{\'\i}tez-Llambay}, P.
  2020, \mnras, 497, 2715, \dodoi{10.1093/mnras/staa1854}

\bibitem[{{Li} \& {Youdin}(2021)}]{li21}
{Li}, R., \& {Youdin}, A.~N. 2021, \apj, 919, 107,
  \dodoi{10.3847/1538-4357/ac0e9f}

\bibitem[{{Lin}(2021)}]{lin21}
{Lin}, M.-K. 2021, \apj, 907, 64, \dodoi{10.3847/1538-4357/abcd9b}

\bibitem[{Ling {et~al.}(2016)Ling, Kurzawski, \& Templeton}]{ling2016reynolds}
Ling, J., Kurzawski, A., \& Templeton, J. 2016, Journal of Fluid Mechanics,
  807, 155

\bibitem[{Long {et~al.}(2015)Long, Shelhamer, \& Darrell}]{long2015fully}
Long, J., Shelhamer, E., \& Darrell, T. 2015, in Proceedings of the IEEE
  conference on computer vision and pattern recognition, 3431--3440

\bibitem[{Maddox {et~al.}(2019)Maddox, Izmailov, Garipov, Vetrov, \&
  Wilson}]{maddox2019simple}
Maddox, W.~J., Izmailov, P., Garipov, T., Vetrov, D.~P., \& Wilson, A.~G. 2019,
  Advances in Neural Information Processing Systems, 32

\bibitem[{{Manger} \& {Klahr}(2018)}]{manger18}
{Manger}, N., \& {Klahr}, H. 2018, \mnras, 480, 2125,
  \dodoi{10.1093/mnras/sty1909}

\bibitem[{Milletari {et~al.}(2016)Milletari, Navab, \& Ahmadi}]{milletari2016v}
Milletari, F., Navab, N., \& Ahmadi, S.-A. 2016, in 2016 fourth international
  conference on 3D vision (3DV), IEEE, 565--571

\bibitem[{{Ormel} \& {Cuzzi}(2007)}]{ormel07}
{Ormel}, C.~W., \& {Cuzzi}, J.~N. 2007, \aap, 466, 413,
  \dodoi{10.1051/0004-6361:20066899}

\bibitem[{Paardekooper {et~al.}(2020)Paardekooper, McNally, \&
  Lovascio}]{paardekooper20}
Paardekooper, S.-J., McNally, C.~P., \& Lovascio, F. 2020, Monthly Notices of
  the Royal Astronomical Society, 499, 4223

\bibitem[{{Pan} \& {Padoan}(2010)}]{pan10}
{Pan}, L., \& {Padoan}, P. 2010, Journal of Fluid Mechanics, 661, 73,
  \dodoi{10.1017/S0022112010002855}

\bibitem[{Pan \& Padoan(2013)}]{pan2013turbulence}
Pan, L., \& Padoan, P. 2013, The Astrophysical Journal, 776, 12

\bibitem[{{Pan} {et~al.}(2011){Pan}, {Padoan}, {Scalo}, {Kritsuk}, \&
  {Norman}}]{pan11}
{Pan}, L., {Padoan}, P., {Scalo}, J., {Kritsuk}, A.~G., \& {Norman}, M.~L.
  2011, \apj, 740, 6, \dodoi{10.1088/0004-637X/740/1/6}

\bibitem[{Ronneberger {et~al.}(2015)Ronneberger, Fischer, \&
  Brox}]{ronneberger2015u}
Ronneberger, O., Fischer, P., \& Brox, T. 2015, in International Conference on
  Medical image computing and computer-assisted intervention, Springer,
  234--241

\bibitem[{{Saffman} \& {Turner}(1956)}]{saffman56}
{Saffman}, P.~G., \& {Turner}, J.~S. 1956, Journal of Fluid Mechanics, 1, 16,
  \dodoi{10.1017/S0022112056000020}

\bibitem[{{Sakurai} {et~al.}(2021){Sakurai}, {Ishihara}, {Furuya}, {Umemura},
  \& {Shiraishi}}]{sakurai21}
{Sakurai}, Y., {Ishihara}, T., {Furuya}, H., {Umemura}, M., \& {Shiraishi}, K.
  2021, \apj, 911, 140, \dodoi{10.3847/1538-4357/abe9ba}

\bibitem[{Sch{\"a}fer {et~al.}(2017)Sch{\"a}fer, Yang, \& Johansen}]{schafer17}
Sch{\"a}fer, U., Yang, C.-C., \& Johansen, A. 2017, Astronomy \& Astrophysics,
  597, A69

\bibitem[{Schaurecker {et~al.}(2021)Schaurecker, Li, Tinker, Ho, \&
  Refregier}]{schaurecker2021super}
Schaurecker, D., Li, Y., Tinker, J., Ho, S., \& Refregier, A. 2021, arXiv
  preprint arXiv:2111.06393

\bibitem[{Shaw(2003)}]{shaw2003particle}
Shaw, R.~A. 2003, Annual Review of Fluid Mechanics, 35, 183

\bibitem[{{Simon} {et~al.}(2016){Simon}, {Armitage}, {Li}, \&
  {Youdin}}]{simon16}
{Simon}, J.~B., {Armitage}, P.~J., {Li}, R., \& {Youdin}, A.~N. 2016, \apj,
  822, 55, \dodoi{10.3847/0004-637X/822/1/55}

\bibitem[{{Squire} \& {Hopkins}(2018)}]{squire18}
{Squire}, J., \& {Hopkins}, P.~F. 2018, \mnras, 477, 5011,
  \dodoi{10.1093/mnras/sty854}

\bibitem[{{Stone} {et~al.}(2020){Stone}, {Tomida}, {White}, \&
  {Felker}}]{stone20}
{Stone}, J.~M., {Tomida}, K., {White}, C.~J., \& {Felker}, K.~G. 2020, \apjs,
  249, 4, \dodoi{10.3847/1538-4365/ab929b}

\bibitem[{Sundaram \& Collins(1997)}]{sundaram_collins_1997}
Sundaram, S., \& Collins, L.~R. 1997, Journal of Fluid Mechanics, 335,
  75–109, \dodoi{10.1017/S0022112096004454}

\bibitem[{{Toschi} \& {Bodenschatz}(2009)}]{toschi09}
{Toschi}, F., \& {Bodenschatz}, E. 2009, Annual Review of Fluid Mechanics, 41,
  375, \dodoi{10.1146/annurev.fluid.010908.165210}

\bibitem[{{van der Marel} {et~al.}(2013){van der Marel}, {van Dishoeck},
  {Bruderer}, {Birnstiel}, {Pinilla}, {Dullemond}, {van Kempen}, {Schmalzl},
  {Brown}, {Herczeg}, {Mathews}, \& {Geers}}]{vdm13}
{van der Marel}, N., {van Dishoeck}, E.~F., {Bruderer}, S., {et~al.} 2013,
  Science, 340, 1199, \dodoi{10.1126/science.1236770}

\bibitem[{{Villaescusa-Navarro} {et~al.}(2021){Villaescusa-Navarro},
  {Angl{\'e}s-Alc{\'a}zar}, {Genel}, {Spergel}, {Somerville}, {Dave},
  {Pillepich}, {Hernquist}, {Nelson}, {Torrey}, {Narayanan}, {Li}, {Philcox},
  {La Torre}, {Maria Delgado}, {Ho}, {Hassan}, {Burkhart}, {Wadekar},
  {Battaglia}, {Contardo}, \& {Bryan}}]{vn21}
{Villaescusa-Navarro}, F., {Angl{\'e}s-Alc{\'a}zar}, D., {Genel}, S., {et~al.}
  2021, \apj, 915, 71, \dodoi{10.3847/1538-4357/abf7ba}

\bibitem[{{Voelk} {et~al.}(1980){Voelk}, {Jones}, {Morfill}, \&
  {Roeser}}]{volk80}
{Voelk}, H.~J., {Jones}, F.~C., {Morfill}, G.~E., \& {Roeser}, S. 1980, \aap,
  85, 316

\bibitem[{Wang {et~al.}(2017)Wang, Wu, \& Xiao}]{wang2017physics}
Wang, J.-X., Wu, J.-L., \& Xiao, H. 2017, Physical Review Fluids, 2, 034603

\bibitem[{{Weidenschilling} \& {Cuzzi}(1993)}]{weidenschilling93}
{Weidenschilling}, S.~J., \& {Cuzzi}, J.~N. 1993, in Protostars and Planets
  III, ed. E.~H. {Levy} \& J.~I. {Lunine}, 1031

\bibitem[{{Whipple}(1972)}]{whipple72}
{Whipple}, F.~L. 1972, in From Plasma to Planet, ed. A.~{Elvius}, 211

\bibitem[{Wu {et~al.}(2022)Wu, Zhao, Shi, \& Chen}]{wu2022large}
Wu, Q., Zhao, Y., Shi, Y., \& Chen, S. 2022, Physics of Fluids, 34, 065129

\bibitem[{Xie {et~al.}(2019)Xie, Li, Ma, \& Wang}]{xie2019modeling}
Xie, C., Li, K., Ma, C., \& Wang, J. 2019, Physical Review Fluids, 4, 104605

\bibitem[{Yang \& Johansen(2016)}]{yang16}
Yang, C.-C., \& Johansen, A. 2016, The Astrophysical Journal Supplement Series,
  224, 39

\bibitem[{Yang {et~al.}(2017)Yang, Johansen, \& Carrera}]{yang17}
Yang, C.-C., Johansen, A., \& Carrera, D. 2017, Astronomy \& Astrophysics, 606,
  A80

\bibitem[{{Yang} \& {Zhu}(2021)}]{yang21}
{Yang}, C.-C., \& {Zhu}, Z. 2021, \mnras, 508, 5538,
  \dodoi{10.1093/mnras/stab2959}

\bibitem[{Yip {et~al.}(2019)Yip, Zhang, Wang, Zhang, Sun, Contardo,
  Villaescusa-Navarro, He, Genel, \& Ho}]{yip2019dark}
Yip, J.~H., Zhang, X., Wang, Y., {et~al.} 2019, arXiv preprint arXiv:1910.07813

\bibitem[{{Youdin} \& {Goodman}(2005)}]{youdin05}
{Youdin}, A.~N., \& {Goodman}, J. 2005, \apj, 620, 459, \dodoi{10.1086/426895}

\bibitem[{{Youdin} \& {Lithwick}(2007)}]{youdin07}
{Youdin}, A.~N., \& {Lithwick}, Y. 2007, \icarus, 192, 588,
  \dodoi{10.1016/j.icarus.2007.07.012}

\bibitem[{Zhu \& Laptev(2017)}]{zhu2017deep}
Zhu, L., \& Laptev, N. 2017, in 2017 IEEE International Conference on Data
  Mining Workshops (ICDMW), IEEE, 103--110

\end{thebibliography}

\appendix
\section{Grid statistics and particle statistics}
\label{sec:grid_stats_vs_part_stats}
Turbulence statistics in prior work \citep[e.g.][]{ishihara2018dust,pan11,pan2013turbulence} are typically calculated from particle positions and velocities directly. However, because the U-Net works with field data, the target particle density and velocity fields must also be binned on a grid to allow training. Similarly, the statistics (e.g. RDF, relative velocity and radial velocity) calculated from our neural network must also be done on a grid. To ensure consistency, we therefore chose to compute both target and particle statistics on a grid. This introduces averaging of particle quantities that might affect our conclusions (and limit comparison against other studies) . Here we attempt to  examine and quantify such effects.

\begin{figure*}
    \centering
    \includegraphics[width=\textwidth]{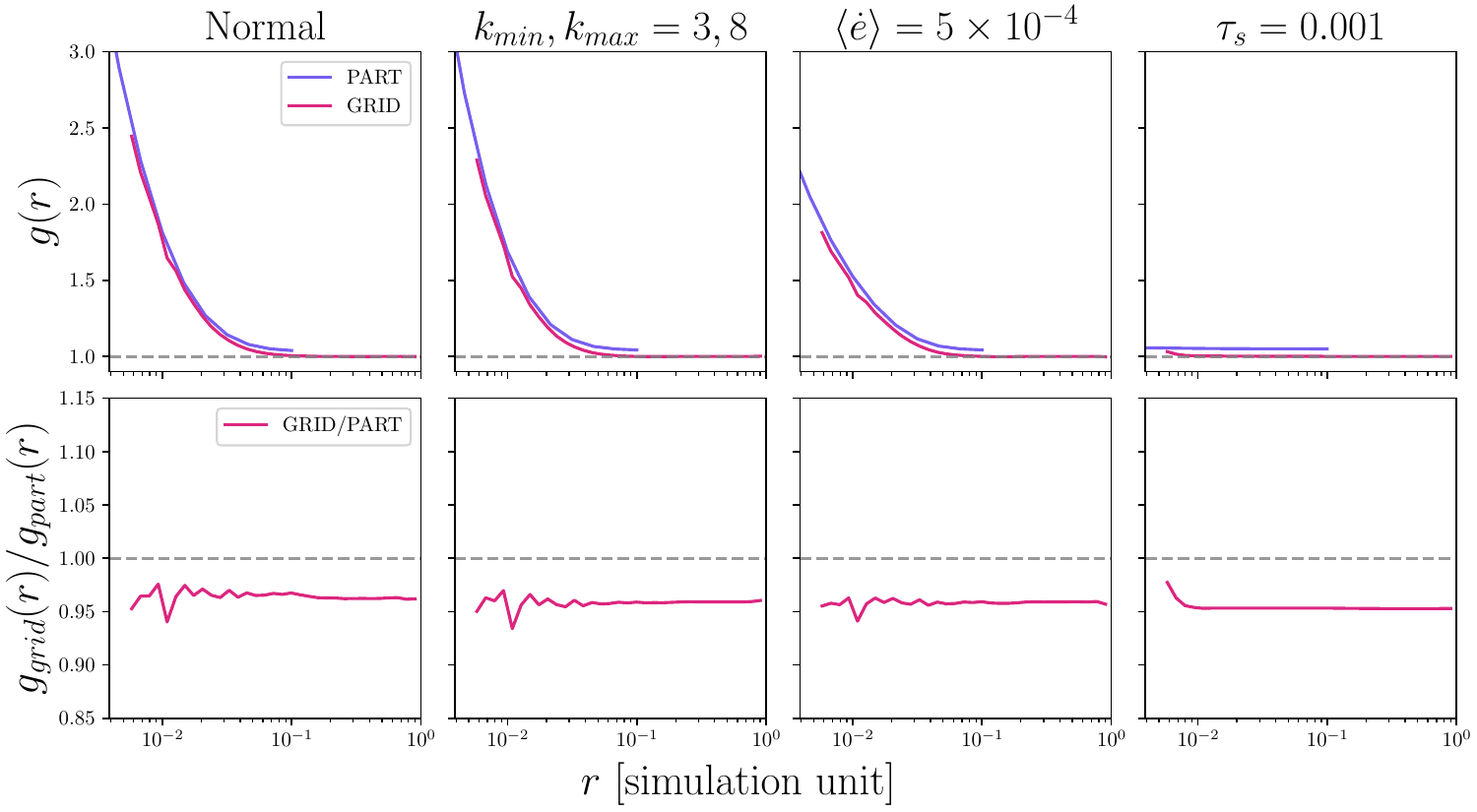}
    \caption{Comparison between grid RDF and particle RDF computed on ground truth data. The particle statistics are computed directly on the simulated particles' positions while the grid result is computed by binning the particles using $256^{3}$ grid points. The analysis is performed for all simulation setups conducted in the paper.}
    \label{fig:grid_vs_part_rdf}
\end{figure*}

In Fig.~\ref{fig:grid_vs_part_rdf} we compare the RDF statistics computed from grid and from particle data. The particle statistics are computed from $r\approx 10^{-5} - 10^{-1}$ while the grid statistics range from $r\approx 10^{-3} - 10^{0}$, limited by grid scale. To aid comparison, the part of the RDF computed for $r < 10^{-3}$ is excluded. The results show that the grid statistics deviate from the particle one for all simulation results by a factor of $\approx 0.9$ at all length scales (the $\tau_s = 0.001$ result is slightly different at the smallest scales, but as we have noted the results for this case are not physically interesting).
%, except for $\tau_s = 0.001$ at $r\lesssim10^{-2}$, where the error appears to be smaller than the rest of the cases by $\approx 5\%$. However, this should not be taken seriously. As discussed in \S\ref{sec:generalize_diff_stop_time}, particles in $\tau_s = 0.001$ cluster at a scale much smaller than the grid scale. Since the entire domain of which the RDF is calculated $\gg \ell_{\tau_s}$, the RDF will only converge to it's asymptotic value 1. Also, because almost all the clustered dust are binned into a single grid point, a sharp increase in RDF at the $r\approx r_{grid}$ is expected. This explains the sudden raise of RDF value in the $\tau_s = 0.001$ plot in Fig.~\ref{fig:grid_vs_part_rdf}.
This offset is likely of systematic nature and the RDFs are likely to be scaled down by the same factor for ML outputs. Therefore, the ratio plots of target and machine learning outputs are likely to be unaffected by the offset because they are both shifted by the same factor. 

\begin{figure}
    \centering
    \includegraphics[width=\textwidth]{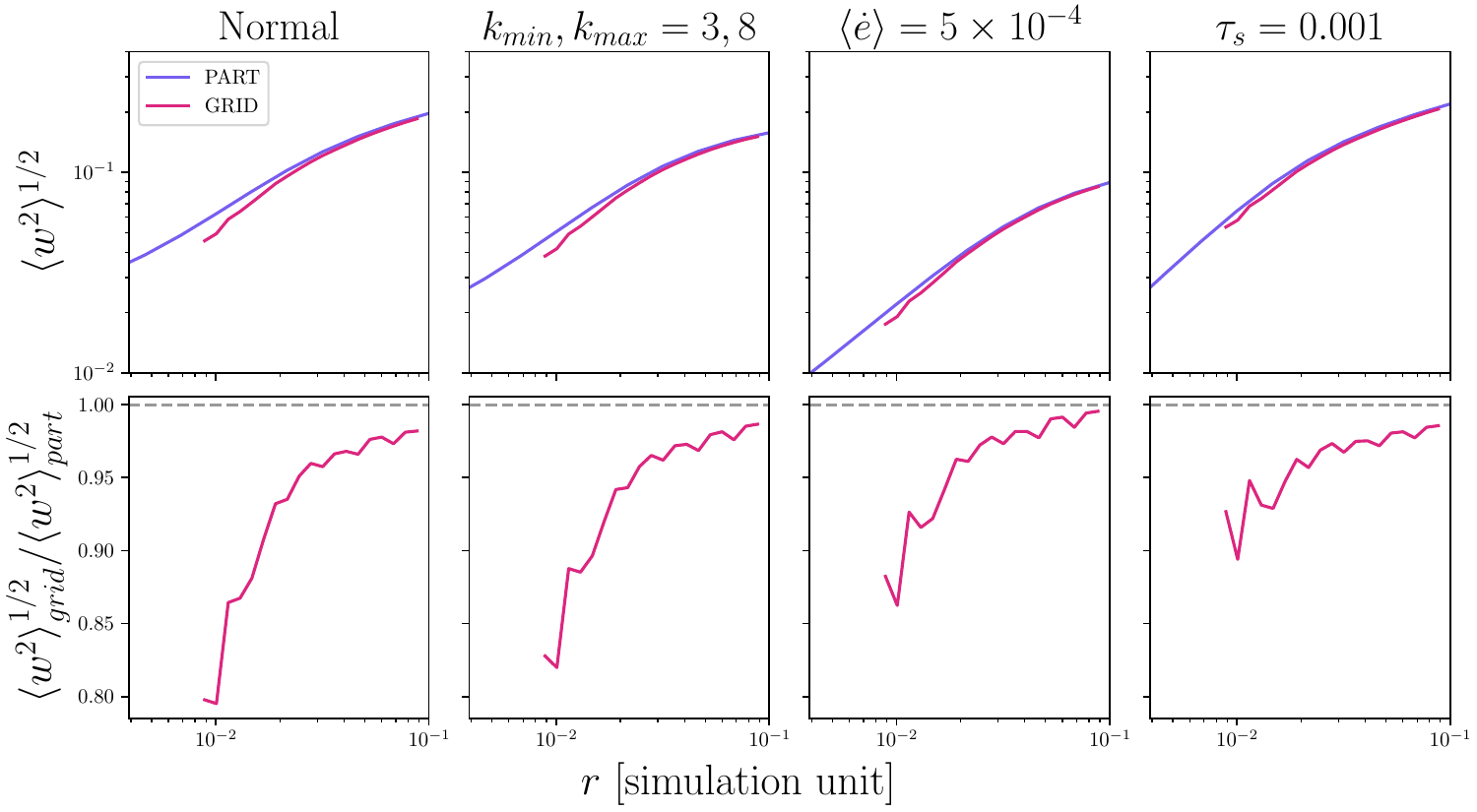}
    \caption{Comparison between relative velocities computed from a grid and relative velocities computed directly from particle data, using the ground truth simulation output.}
    \label{fig:grid_vs_part_rel_vel}
\end{figure}

Similarly, we compare the relative velocities computed by the two methods for all test cases and summarize them in Fig.~\ref{fig:grid_vs_part_rel_vel}. Like the RDFs, the grid relative velocities have a similar shape to the particle ones and are below the particle ones at all lengths. However, unlike the RDFs, the deviation shows a radial dependence. The deviation increases gradually from $\approx 2.5\%$ to $7.5\%$ from $10^{-1}\gtrsim r \gtrsim 2\times 10^{-2}$, after which it increases rapidly to at most $20\%$ at a radial interval of $\approx 10^{-2}$. Before the plunge at small scales, the {\em difference} between the deviations of all test cases is bounded by an interval of $\approx 2.5\%$. This suggests that the relative velocities computed on those scales are relatively stable and consistent. However, at small scales, the deviations between different test cases spread out to $\approx 12.5\%$. We interpret the difference between the grid and particle-computed statistics as being a consequence of reducing the true particle data---which is not single valued at a point in space---to a gridded vector field. This reduction effectively throws away a term related to the velocity dispersion of the particles at the grid scale, introducing a scale-dependent error.

%The discrepancy between particle and grid statistics can be understood in the following way: When the scale is large, the dust flow in a more coherent matter. Therefore, the error due to the averaging effects induced in the gridding process is small. However, the dispersion in velocities increases as smaller $r$. In binning the velocities in grid, the velocities are averaged into smaller values. This make the grid calculated version smaller than the one computed directly on the particles. 

These tests suggest that the velocity statistics computed from the grid are reliable for scales $r\gtrsim 2\times 10^{-2} $. At smaller scales the gridded representation differs systematically from that computed directly from particle data. This offset is of limited significance for our study---where we have consistently made all comparisons on the grid---but would need to be considered carefully if a model similar to ours was used to draw conclusions near grid scales.

\section{Consistency across frames}
\label{sec:consistency_across_frames}

\begin{figure*}
    \centering
    \includegraphics[width=0.95\textwidth]{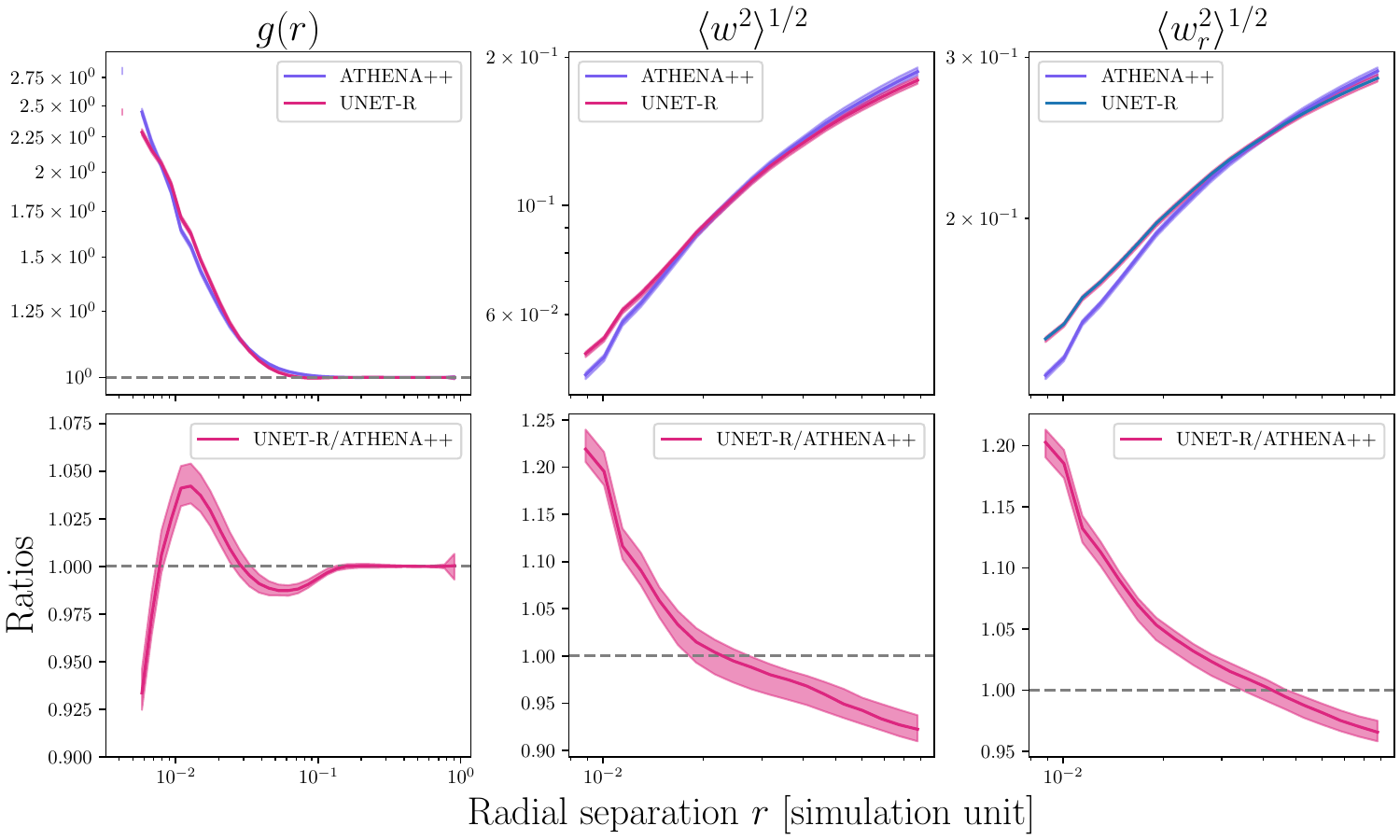}
    \caption{Consistency of particle statistics across frames. We compute the statistics for the entire validation set and plot the median as the solid lines. The lower and upper edge of the color band represents the 5th percentile and 95th percentile of the distribution. Therefore, $90\%$ of the values computed across the validation set lie  within the band.}
    \label{fig:normal_consistency}
\end{figure*}

The plots shown in the main body of the paper are computed from a single, arbitrarily chosen, frame. In Fig.~\ref{fig:normal_consistency}, we compute the statistics on the entire validation set and quantify the frame-to-frame variation. From the figure, we see that the color band representing the absolute fluctuation between the 5th percentile and 95th percentile is barely observable. Also, the fluctuation in the ratio plots is limited to a $\lesssim3\%$ band centered at the median (solid line) and does not affect the general shape of the curve obtained. We therefore believe that our discussion above generalizes across frames. 

\section{Error estimation of statistics}
\label{sec_errors}
Estimating the uncertainty in model predictions is essential for extracting reliable information from machine learning predictions. Roughly speaking, the uncertainty of a network can be broadly classified as model uncertainty and data uncertainty \citep{kendall2017uncertainties}. The former quantifies our uncertainty of the predictive model, and the latter quantifies our uncertainty of the input data. Because our data are simulated, we will only concern ourselves with model uncertainties. 

Ideally, model uncertainties can be quantified by placing a distribution over model parameters and marginalizing to give a predictive distribution. However, in modern neural networks this is largely intractable \citep{kendall2017uncertainties,maddox2019simple}. Therefore, various approximation methods have been developed by the machine learning community to approximate the procedure. We chose to adopt and compare two of the most popular frameworks: Monte-Carlo Dropout (MC Dropout) \citep{gal2016dropout,zhu2017deep} and SWA-Gaussian (SWAG) \citep{maddox2019simple} to quantify our errors. 

\subsection{Monte-Carlo Dropout (MC Dropout)}
In MC dropout, dropout layers are added before every weight layer with a dropout probability $p$. Layers with dropout drop weights randomly with a probability of $p$. \cite{gal2016dropout} has shown that by dropping out at  validation time and averaging out the results, we can approximate the first moment of the predicted distribution as,
\begin{equation}
    \hat{\mathbf{y}}_{MC} = \frac{1}{T}\sum_{i=1}^{T} \hat{\mathbf{y}}_i(\hat{\mathbf{x}}),
    \label{eq:dropout_mean}
\end{equation}
where $\hat{\mathbf{y}}_i$ is the output of the $i$-th validation run using the input $\hat{\mathbf{x}}$ and $T$ is the number of forward passes used during validation. For the second moment, we compute the variance of the predicted  distribution as,

\begin{equation}
    \sigma_\text{tot}^2 = \sigma_\text{model}^2 + \sigma_\text{inherent}^2,
    \label{eq:dropout_uncertainty}
\end{equation}
where $\sigma_\text{model}$ is the model uncertainty and $\sigma_\text{inherent}$ is the data noise uncertainty. Since we want to focus on model uncertainty, we set $\sigma_\text{inherent}$ to zero, 
\begin{align}
    \sigma_\text{model}^2 &= \frac{1}{T} \sum_{i=1}^{T} \left(\hat{\mathbf{y}}_i - \hat{\mathbf{y}}_{MC}\right)^2,
\end{align}
where $\hat{\mathbf{y}}_i$ and $\hat{\mathbf{y}}_\text{MC}$ take the same meaning as above. We trained 4 networks with dropout rates $p=0.1,0.3,0.5,0.7$ and sampled 100 times for each network. The particle statistics of the generated frames are presented in Fig.~\ref{fig:swag_vs_mc_dropout_stats}. 
We find, first, that the predicted mean of the dropout networks becomes less accurate when $p$ increase. Second, the model fluctuations, represented by the color bars in the plot, are too small to be observed even with $p=0.7$. This indicates that the results of the model are precise but not accurate. 

\subsection{SWA-Gaussian (SWAG)}

We also performed sampling using another popular method: SWA-Gaussian \citep{maddox2019simple}. In the method, we sample the model parameters near to those of the pre-trained network using a high constant learning rate. The collected model weights are then averaged to get the average weight $\theta_\text{SWA}$ using a simple average. The low rank and diagonal approximation of the weights' covariance $\Sigma_\text{low},\Sigma_\text{diag}$ are also computed. This results in a Gaussian approximation of the posterior over weight parameters, which can be sampled to perform a Bayesian model average. In our case, we trained our SWAG network using a constant learning rate of $\text{lr}=0.02$ for a few hundred epochs. The extracted $\theta_\text{SWA},\Sigma_\text{low},\Sigma_\text{diag}$ are then used to generate 20 samples, which are used to give a model averaged RDF, relative velocities and their error bars. 

\begin{figure*}
    \centering
    \includegraphics[width=\textwidth]{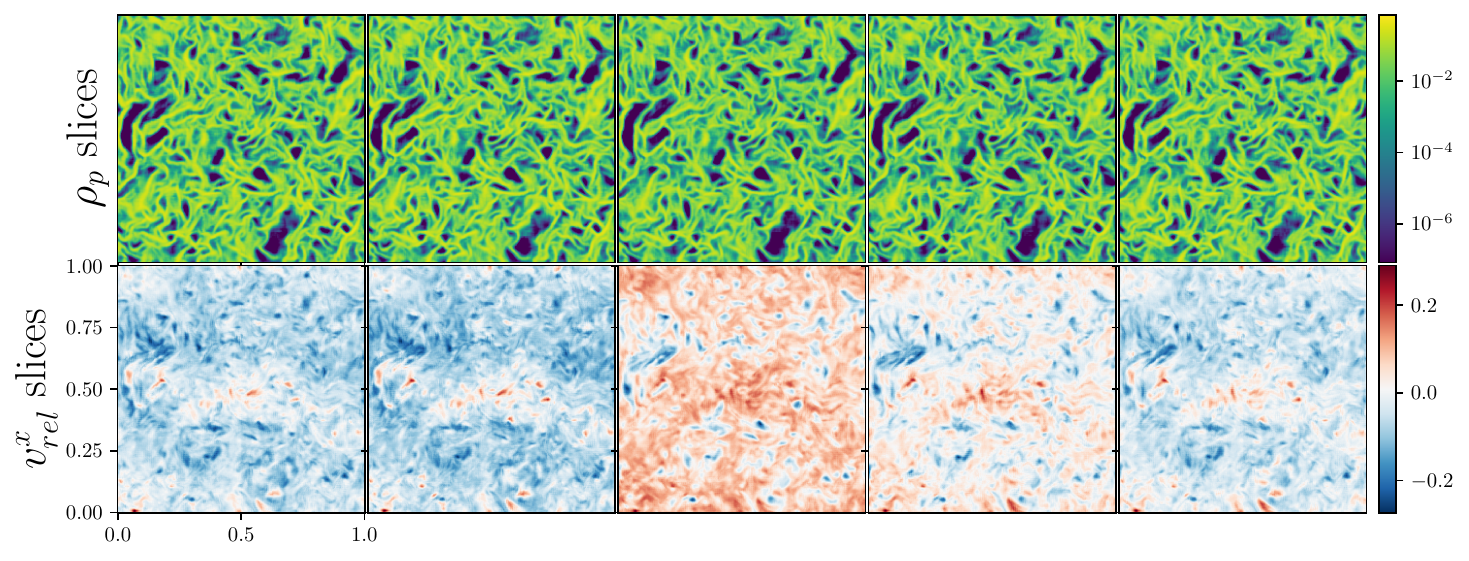}
    \caption{Examples of density fields and velocity fields generated by SWAG. Here we only show 5 of 20 samples generated. }
    \label{fig:swag_slices}
\end{figure*}

We randomly selected 5 samples generated by SWAG and plotted slices of the output particle density $\rho_p$ and  particle relative velocity field along the $x$-direction in Fig.~\ref{fig:swag_slices}. As shown in the figure, all $\rho_p$ fields generated by SWAG are of similar shape and have similar features. This suggests the model uncertainty of $\rho_p$ generated is small. On the other hand, two of the five slices of the output $v_{rel}^{x}$ differ significantly both from the other slices and from the true output field shown in Fig.~\ref{fig:vrel_slice_comp}. This indicates a larger model uncertainty in the generated relative velocity field. 

\begin{figure*}
    \centering
    \includegraphics[width=0.8\textwidth]{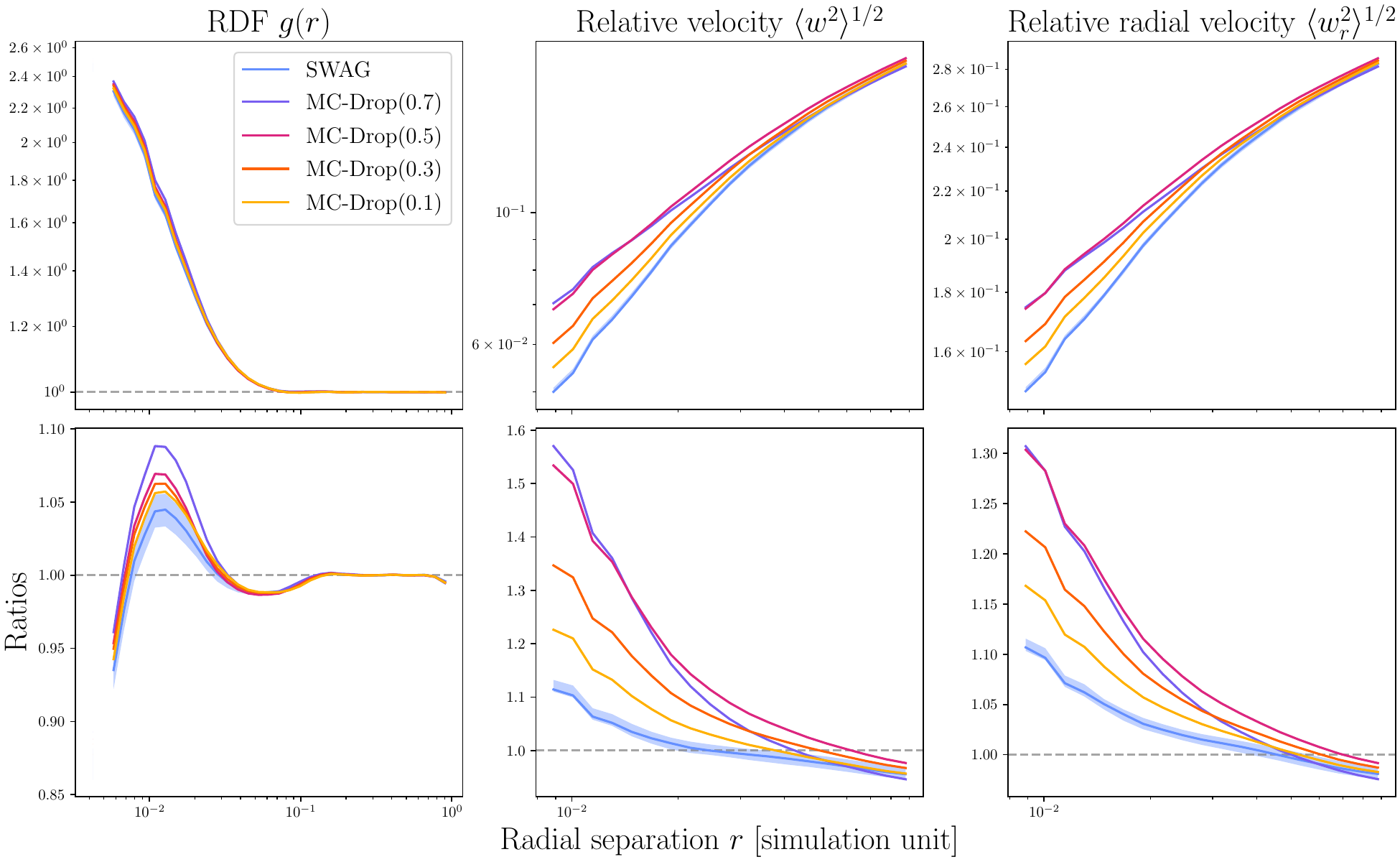}
    \caption{Comparison between SWAG output and MC Dropout outputs with different dropout probabilities. Samples generated by the networks are used to generate the statistics, and the distribution is plotted. The solid lines are the median, and the upper and lower portion of the color bars are the upper and lower quantile of the distribution. Note that the color bars for MC dropouts are all too small to be visible on the graph. }
    \label{fig:swag_vs_mc_dropout_stats}
\end{figure*}

The particle statistics are presented in Fig.~\ref{fig:swag_vs_mc_dropout_stats}. From the figure, we see that the fluctuation predicted by SWAG is much larger than MC dropout. Hence, the error bars predicted by the two frameworks are not consistent. In agreement with what we see visually in output slices, we find that the uncertainty in velocity statistics as estimated using SWAG is much larger than that of particle density $\rho_p$.  However, even with larger error bars, none of the error bands are large enough to mask the discrepancy between the prediction and the output. Therefore, we conclude that the discrepancy between ML predictions and target is systematic in both SWAG and MC-Dropout.  

In all statistics, the errors of the predicted mean using MC dropout are always larger than that of SWAG. This is expected, because by dropping out parameters randomly during training and validation, we effectively average a fraction of parameters in the model to take similar values. This should reduce the effective model size of MC-Dropout and reduce its expressive power.

\end{CJK*}
\end{document}